\documentstyle[aps,epsf,epsfig]{revtex}

\newcommand{\bq}{\begin{equation}}
\newcommand{\eq}{\end{equation}}
\newcommand{\bqn}{\begin{eqnarray}}
\newcommand{\eqn}{\end{eqnarray}}
\newcommand{\nb}{\nonumber}
\newcommand{\lb}{\label}
\begin{document}

\title{Topological Charged Black Holes in High Dimensional Spacetimes
and Their Formation from Gravitational Collapse of a
Type II Fluid}
\author{Yumei Wu \thanks{E-mail: yumei@dmm.im.ufrj.br} $^{1}$,
M. F. A. da Silva\thanks{E-mail: mfas@dtf.if.uerj.br} $^{2}$, N.
O. Santos\thanks{E-mail: nos@cbpf.br} $^{3,4}$, and Anzhong
Wang\thanks{E-mail: Anzhong$\_$Wang@baylor.edu} $^{2,5,6}$}
\address{$^{1}$ Instituto de Matem\'atica, Universidade Federal do Rio
de Janeiro, Caixa Postal $68530$, CEP. $21945-970$,
Rio de Janeiro~--~RJ, Brazil\\
$^{2}$ Departamento de F\' {\i}sica Te\' orica,
Universidade do Estado do Rio de Janeiro,
Rua S\~ ao Francisco Xavier 524, Maracan\~a,
CEP. $20550-013$, Rio de Janeiro~--~RJ, Brazil\\
$^3$ Laborat\'orio Nacional de Computa\c{c}\~ao Cient\'{\i}fica,
CEP. $25651-070$, Petr\'opolis-RJ, Brazil \\
 $^4$  Centro Brasileiro de Pesquisas F\'{\i}sicas, CEP. $22290-180$,
Rio de Janeiro-RJ, Brazil \\
$^5$ CASPER, Physics Department, P.O. Box 97316, Baylor
University,  Waco, TX76798-7316\\
$^6$ Department of Physics, University of Illinois at
Urbana-Champaign, 1110 West Green Street, Urbana, IL 61801-3080}


\date{\today}

\maketitle

\begin{abstract}

Topological charged black holes coupled with a cosmological constant in
$R^{2}\times X^{D-2}$  spacetimes are studied, where $X^{D-2}$ is an
Einstein space of the form ${}^{(D-2)}R_{AB} = k(D-3) h_{AB}$. The global
structure for the four-dimensional spacetimes with $k = 0$ is investigated
systematically. The most general solutions that represent a Type $II$ fluid
in such a high dimensional spacetime are found, and showed that topological
charged black holes can be formed from the gravitational collapse of such a
fluid.  When the spacetime is (asymptotically) self-similar, the
collapse always forms black holes for $k = 0, \; -1$, in contrast to
the case $k = 1$, where it can form either balck holes or naked
singularities.

\end{abstract}

\section{Introduction}

Lately black holes in high dimensions have attracted a great deal of attention
in the gravity-gauge theory correspondence \cite{Mal98}, and been further promoted
by theories of TeV gravity, in which high dimensional black holes are predicted
to be produced in the next generation of colliders \cite{DL01}.

In black hole physics, one of the fundamental features is the topology of
a black hole. In four-dimensional asymptotically flat stationary spacetimes, Hawking
first showed that a black hole has necessarily   a $S^{2}$ topology, provided
that the dominant energy condition holds \cite{Haw72}. Later, it was realized that
Hawking's theorem can be  improved in various aspects, see
\cite{CG01} and references therein. However, once the energy condition is relaxed,
a black hole can have quite different topologies. Such examples
can ocurr even in $3+1$ dimensional spacetimes where the cosmological constant is
negative \cite{TBHs1}. In high dimensional spacetimes, it was found that even
if the energy conditions hold, the topology is still not unique. In particular, a
five-dimensional rotating vacuum black hole can have either a $S^{3}$ topology
\cite{MP86}, or a $S^{2}\times S^{1}$ topology \cite{ER02}, although
a static charged dilaton black hole in high dimensional asymptotically flat
spacetimes indeed has an unique topology, which is  a $(D-2)$-sphere \cite{Gib02}.
When the cosmological constant is different from zero, similar to the
four-dimensional case,  static black holes can have  different topologies
\cite{TBHs2}.

In this paper we shall study the formation of topological black holes
from gravitational collapse of a type $II$ fluid in high dimensional
spacetimes. Specifically, the paper is organized as follows: In Sec. $II$
we present a general $R^{2}\times X^{D-2}$ decomposition.
In Sec. $III$, assuming that $X^{D-2}$ is an Einstein space with a constant curvature,
$^{(D-2)}R_{AB} = k_{D}h_{AB}$, we re-derive the charged solutions
coupled with a cosmological constant in any dimensional spacetimes,
without  assuming  that the spacetime is static. In Sec. $IV$ we systematically
study the global structure for  the case $D = 4$ and $k_{D} = 0$, while in
Sec. $V$, all the type $II$ fluid solutions in D-dimensional spacetimes are given.
In Sec. $VI$ we study the formation of topological charged and uncharged
 black holes from the
gravitational collapse of such a fluid, while in Sec. $VII$ our main conclusions
are summarized. There is also an appendix, in which trapped surfaces
and apparent horizons are defined.

Before proceeding further, we would like to note that the formation of
topological black holes from gravitational collapse
in four-dimensional spacetimes was studied in \cite{SM97}, while gravitational
collapse in high dimensional spherically symmetric spacetimes  was
investigated in \cite{Ban94}.

\section{$R^{2}\times X^{D-2}$ Decomposition}
\lb{SecII}
\renewcommand{\theequation}{2.\arabic{equation}}
\setcounter{equation}{0}

In this paper let us  consider a D-dimensional spacetime described by the metric
\bq
\lb{2.1}
ds^{2} = g_{\mu\nu} dx^{\mu}dx^{\nu} = \gamma_{ab}(x^{c})dx^{a}dx^{b}
+ s^{2}(x^{c})h_{AB}(x^{C})dx^{A}dx^{B},
\eq
where we use Greek indices, such as, $\mu,\; \nu,\; \lambda$, to run
from $0$ to $D-1$,  lowercase Latin indices, such as, $a,\; b,\; c,\; ...$, to run
from $0$ to $1$, and uppercase Latin indices,   such as, $A,\; B,\; C ...$, to run
from $2$ to $D-1$. Clearly, the above metric is invariant
under the coordinate transformations,
\bq
\lb{2.1a}
x^{a} = x^{a}({x'}^{b}),\;\;\;
x^{A} = x^{A}({x'}^{B}).
\eq
Introducing the quantities $\gamma^{ab}(x^{c})$ and
$h^{AB}(x^{C})$ via the relations,
\bq
\lb{2.2}
\gamma^{ac}\gamma_{cb} = \delta^{a}_{b},\;\;\;\;
h^{AC}h_{CB} = \delta^{A}_{B},
\eq
we find that
\bq
\lb{2.3}
g_{\mu\nu} = \left(\matrix{\gamma_{ab} & 0\cr
0 & s^{2}h_{AB}\cr}\right),\;\;\;\;
g^{\mu\nu} = \left(\matrix{\gamma^{ab} & 0\cr
0 & s^{-2}h^{AB}\cr}\right).
\eq
Then, the non-vanishing Christoffel symbols are given by
\bqn
\lb{2.5}
{}^{(D)}\Gamma^{a}_{bc} &=& {}^{(2)}\Gamma^{a}_{bc},\;\;\;
{}^{(D)}\Gamma^{a}_{AB} = -  h_{AB} s  s^{,a},\nb\\
{}^{(D)}\Gamma^{A}_{aB} &=& s^{-1}{s_{,a}} \delta^{A}_{B},\;\;\;
{}^{(D)}\Gamma^{A}_{BC} = {}^{(n)}\Gamma^{A}_{BC},
\eqn
where $s_{,a} \equiv \partial s/\partial x^{a},\; s^{,a} \equiv \gamma^{ab} s_{,b},\;
{}^{(2)}\Gamma^{a}_{bc}$ and ${}^{(n)}\Gamma^{A}_{BC}$
are the Christoffel symbols calculated, respectively, from
$\gamma_{ab}(x^{c})$ and $h^{AB}(x^{C})$, and $n \equiv D -2$.
From Eq.(\ref{2.5}) we find that  the
Ricci tensor, defined by,
\bq
\lb{2.6}
{ }^{(D)}R_{\mu\nu} = { }^{(D)}\Gamma^{\lambda}_{\mu\nu,\lambda}
- { }^{(D)}\Gamma^{\lambda}_{\lambda\mu,\nu}
+ { }^{(D)}\Gamma^{\lambda}_{\lambda\sigma}
  { }^{(D)}\Gamma^{\sigma}_{\mu\nu}
- { }^{(D)}\Gamma^{\lambda}_{\sigma\nu}
  { }^{(D)}\Gamma^{\sigma}_{\lambda\mu},
\eq
has the   following non-vanishing components,
\bqn
\lb{2.7}
{ }^{(D)}R_{ab}  & =& { }^{(2)}R_{ab}
- \frac{D-2}{s}  \nabla_{a}\nabla_{b}s,\nb\\
{ }^{(D)}R_{AB}  & =& { }^{(n)}R_{AB} - \left[s\Box{s}
+ \left(D-3\right) \left(\nabla_{c}s\right)
\left(\nabla^{c}s\right)\right] h_{AB},
\eqn
where $\Box \equiv \gamma^{ab}\nabla_{a}\nabla_{b}$, and $\nabla_{a}$
denotes the covariant derivative with respect to $\gamma_{ab}$.

To calculate ${ }^{(2)}R_{ab}$, we first note that the Riemann tensor
${ }^{(2)}R_{abcd}$ has only one independent component, say,
${ }^{(2)}R_{0101}$. Then, in terms of ${ }^{(2)}R_{0101}$ we have
\bqn
\lb{2.9}
{ }^{(2)}R_{abcd} &=& { }^{(2)}R_{0101}
\left(\delta^{0}_{a}\delta^{1}_{b}\delta^{0}_{c}\delta^{1}_{d}
- \delta^{0}_{a}\delta^{1}_{b}\delta^{1}_{c}\delta^{0}_{d}
- \delta^{1}_{a}\delta^{0}_{b}\delta^{0}_{c}\delta^{1}_{d}
+ \delta^{1}_{a}\delta^{0}_{b}\delta^{1}_{c}\delta^{0}_{d}\right),\nb\\
{ }^{(2)}R_{ab} &\equiv& { }^{(2)}{R^{c}}_{acb}
= { }^{(2)}R_{0101} \left(\gamma^{11}\delta^{0}_{a}\delta^{0}_{b}
- \gamma^{01}\left(\delta^{0}_{a}\delta^{1}_{b}
+ \delta^{1}_{a}\delta^{0}_{b}\right)
+ \gamma^{00}\delta^{1}_{a}\delta^{1}_{b}\right).
\eqn

On the other hand, for the second block of the metric,  we shall
consider only the case where $h_{AB}$ represents a $n$-dimensional surface
with constant curvature,
\bq
\lb{2.8}
{ }^{(n)}R_{AB} = k_{D} h_{AB}(x^{C}),
\eq
where, without loss of generality, one can always set $k_{D} = (D-3)k$
with $k = -1, \; 0, \; 1$. Then the surfaces of constant $x^{a}$  will be referred,
respectively, to as {\em hyperbolic, flat and elliptic}.

\section{ Topological Charged Black Holes Coupled with the Cosmological Constant}
\lb{SecIII}
\renewcommand{\theequation}{3.\arabic{equation}}
\setcounter{equation}{0}

Using the gauge freedom of Eq.(\ref{2.1a}), in this section we shall choose the
coordinates such that
\bq
\lb{3.1}
\gamma_{t r}(x^{c}) = 0,\;\;\; s(x^{c}) = r,
\eq
where $x^{c} = \{t, r\}$. This will be referred to as {\em the Schwarzschild gauge}.
Then, we find that
\bq
\lb{3.2}
ds^{2} =  - e^{2\Phi(t,r)}dt^{2} + e^{2\Psi(t,r)}dr^{2} +
r^{2}h_{AB}(x^{C})dx^{A}dx^{B}.
\eq
For such a form, there still remains the gauge freedom
\bq
\lb{3.2a}
t = t(\bar{t}).
\eq
Later, we shall use it to fix some integration functions.
For the metric (\ref{3.2}) we find
\bqn
\lb{3.4}
{}^{(2)}R_{ab} &=& - \frac{1}{2}\; {}^{(2)}R
\left(e^{2\Phi}\delta^{0}_{a}\delta^{0}_{b}
- e^{2\Psi}\delta^{1}_{a}\delta^{1}_{b}\right),\nb\\
{}^{(2)}R &=& 2\left\{e^{-2\Phi}\left[\Psi_{,tt}
+ \Psi_{,t}\left(\Psi_{,t} -\Phi_{,t}\right)\right]
 - e^{-2\Psi}\left[\Phi_{,rr}
+ \Phi_{,r}\left(\Phi_{,r} -\Psi_{,r}\right)\right]\right\}.
\eqn
It can be also shown that
\bqn
\lb{3.5}
\nabla_{a}\nabla_{b}r &=& - e^{2(\Phi -\Psi)}\Phi_{,r}
\delta^{0}_{a}\delta^{0}_{b}
- \Psi_{,t}\left(\delta^{0}_{a}\delta^{1}_{b}
+ \delta^{1}_{a}\delta^{0}_{b}\right)
- \Psi_{,r} \delta^{1}_{a}\delta^{1}_{b},\nb\\
\Box r &=& e^{-2\psi}\left(\Phi_{,r} - \Psi_{,r}\right).
\eqn
Substituting Eqs.(\ref{2.8}), (\ref{3.4}) and (\ref{3.5}) into Eq.(\ref{2.7})
we obtain
\bqn
\lb{3.6a}
{}^{(D)}R_{00} &=& - \left[\Psi_{,tt}
+ \Psi_{,t}\left(\Psi_{,t} - \Phi_{,t}\right)\right]\nb\\
& & + e^{2(\Phi - \Psi)}\left[\Phi_{,rr}
+ \Phi_{,r}\left(\Phi_{,r} - \Psi_{,r} + \frac{D-2}{r}\right)\right],\\
\lb{3.6b}
{}^{(D)}R_{01} &=& \frac{D-2}{r}\Psi_{,t},\\
\lb{3.6c}
{}^{(D)}R_{11} &=& e^{2(\Psi - \Phi)}\left[\Psi_{,tt}
+ \Psi_{,t}\left(\Psi_{,t} - \Phi_{,t}\right)\right]\nb\\
& & - \left(\Phi_{,rr}
+ \Phi_{,r}\left(\Phi_{,r} - \Psi_{,r}\right)
-  \frac{D-2}{r} \Psi_{,r}\right),\\
\lb{3.6d}
{}^{(D)}R_{AB} &=& \left\{k_{D} -  \left[r \left(\Phi_{,r} - \Psi_{,r}\right)
  + \left(D-3\right)\right]e^{-2\Psi}\right\}h_{AB}.
\eqn

On the other hand, the D-dimensional Einstein-Maxwell equations with
the cosmological constant, $\Lambda_{D}$,   read
\bq
\lb{3.7}
{}^{(D)}R_{\mu\nu} - \frac{1}{2}g_{\mu\nu}{}^{(D)}R = 8\pi G_{D}\;
{}^{(D)}E_{\mu\nu} - \Lambda_{D}g_{\mu\nu},
\eq
where the energy-momentum tensor ${}^{(D)}E_{\mu\nu}$ is given by
\bq
\lb{3.8}
{}^{(D)}E_{\mu\nu} = \frac{1}{4\pi
G_{D}}\left({}^{(D)}F_{\mu\alpha}{{}^{(D)}F_{\nu}}^{\alpha}
- \frac{1}{4}g_{\mu\nu} \; {}^{(D)}F_{\alpha\beta}\;
{}^{(D)}F^{\alpha\beta}\right),
\eq
with the electromagnetic field ${}^{(D)}F_{\mu\nu}$ satisfying the Maxwell
equations,
\bq
\lb{3.9}
D^{\nu}\; {}^{(D)}F_{\mu\nu} = 0,
\eq
where $D_{\nu}$ denotes the covariant derivative with respect to the
D-dimensional metric $g_{\mu\nu}$. From the symmetry of the spacetime, without
loss of generality, we can always assume that   ${}^{(D)}F_{\mu\nu}$  is
function of $t$ and $r$ only, and takes the form,
\bq
\lb{3.10}
{}^{(D)}F_{\mu\nu} = F(t, r)\left(\delta^{0}_{\mu}\delta^{1}_{\nu} -
\delta^{1}_{\mu}\delta^{0}_{\nu}\right).
\eq
Substituting Eqs.(\ref{3.8}) and (\ref{3.9}) into Eq.(\ref{3.7}) we find that
\bqn
\lb{3.11}
{}^{(D)}R_{\mu\nu} &=& {F}^{2}\left(2\left(e^{-2\Psi}
\delta^{0}_{\mu}\delta^{0}_{\nu}
- e^{-2\Phi}\delta^{1}_{\mu}\delta^{1}_{\nu}\right)
+ \frac{2}{D-2} e^{-2(\Phi + \Psi)} g_{\mu\nu}\right)\nb\\
& & + \frac{2\Lambda_{D}}{D-2}g_{\mu\nu}.
\eqn
Combining Eq.(\ref{3.6b}) with the above equation, we obtain
\bq
\lb{3.12}
\Psi(t, r) = \Psi(r).
\eq
On the other hand, from the Maxwell field equation (\ref{3.9}) we have
\bq
\lb{3.13}
F(t, r) = \frac{C_{0}}{r^{D-2}}e^{\Phi + \Psi},
\eq
where $C_{0}$ is an integration constant.
Inserting Eqs.(\ref{3.12}) and (\ref{3.13}) into
Eqs.(\ref{3.6a})-(\ref{3.6d}), we find the following independent equations,
\bqn
\lb{3.14a}
\Phi_{,rr} + \Phi_{,r}\left(\Phi_{,r} - \Psi_{,r} + \frac{D-2}{r}\right) &=&
e^{2\Psi}\left(\frac{C_{1}}{r^{2(D-2)}} - \frac{2\Lambda_{D}}{D-2}\right),\\
\lb{3.14b}
\Phi_{,rr} + \Phi_{,r}\left(\Phi_{,r} - \Psi_{,r}\right)-
\frac{D-2}{r}\Psi_{,r} &=& e^{2\Psi}\left(\frac{C_{1}}{r^{2(D-2)}} -
\frac{2\Lambda_{D}}{D-2}\right),\\
\lb{3.14c}
k_{D}e^{2\Psi} - r\left(\Phi_{,r} - \Psi_{,r}\right) - (D-3)  &=&
e^{2\Psi}\left(\frac{C_{1}}{(D-3) r^{2(D-3)}}
 + \frac{2\Lambda_{D}}{D-2}r^{2}\right),
\eqn
where $C_{1} \equiv 2(D-3)C^{2}_{0}/(D-2)$. From Eqs.(\ref{3.14a}) and
(\ref{3.14b}) we obtain
\bq
\lb{3.15}
\Phi(t, r) = - \Psi(r) + \Phi_{0}(t),
\eq
where $\Phi_{0}(t)$ is an arbitrary function. However, using the coordinate
transformation (\ref{3.2a}), we can always set $\Phi_{0}(t) = 0$,
a condition that will be assumed below. Then, substituting the above
expression into Eq.(\ref{3.14c}) and integrating it, we obtain
\bq
\lb{3.16}
\Phi = - \Psi = \frac{1}{2}\ln\left(k - \frac{2M}{r^{D-3}} +
\frac{Q^{2}}{r^{2(D-3)}} - \Lambda r^{2}\right),
\eq
where
\bq
\lb{3.16a}
\Lambda \equiv \frac{2\Lambda_{D}}{(D-1)(D-2)},
\eq
and $M$ and $Q$ are two arbitrary constants, related to the total mass $M_{D}$
and charge $Q_{D}$ of the spacetime via the relations \cite{CS98},
\bq
\lb{3.17}
M = \frac{8\pi G_{D} M_{D}}{(D-2) A_{\Sigma}},\;\;\;
Q = \frac{4\pi G_{D} Q_{D}}{A_{\Sigma}},
\eq
where $A_{\Sigma}$ denotes the total area of the $(D-2)$-dimensional surface
spanned by $h_{AB}$,
\bq
\lb{3.18}
A_{\Sigma} = \int{\sqrt{h} \; d^{D-2}x}.
\eq
It can be shown that the solution
(\ref{3.16}) also satisfies Eqs.(\ref{3.14a}) and (\ref{3.14b}), with the
electromagnetic field given by
\bq
\lb{3.19}
{}^{(D)}F_{\mu\nu} =\left(\frac{(D-2)(D-3)}{2}\right)^{1/2} \frac{Q}{r^{D-2}}
\left(\delta^{0}_{\mu}\delta^{1}_{\nu} -
\delta^{1}_{\mu}\delta^{0}_{\nu}\right).
\eq

When the surfaces of constant $t$ and $r$ are not compact, clearly $A_{\Sigma}$
becomes infinity, so do $M_{D}$ and $Q_{D}$. However, one can still keep
the quantities $m$ and $q$ finite, where
\bq
\lb{3.17a}
m \equiv \frac{M_{D}}{A_{\Sigma}},\;\;\;\;
q \equiv \frac{Q_{D}}{A_{\Sigma}},
\eq
which may be interpreted as the mass and charge per unit area.

On the other hand, from the above derivation of
the solutions one can see that they represent
 the most general solutions of the Einstein-Maxwell
field equations coupled with the cosmological constant in D-dimensional
spacetimes described by the metric (\ref{3.2}). In general the spacetime is singular
at $r = 0$ and all the other singularities are coordinate ones. This can be seen,
for example, from the Ricci scalar, which now is given by
\bq
\lb{3.20}
{}^{(D)}R = - \frac{(D-3)(D-4)Q^{2}}{r^{2(D-2)}} + \frac{2D}{D-2} \Lambda_{D}.
\eq
 Thus, to have a
geodesically complete spacetime one needs to extend the spacetime beyond these
coordinate singularities. In the next section, we shall restrict ourselves
only to the case where $k_{D} =  0$ and $D = 4$.

\section{ Flat Topological Black Holes in Four-Dimensional Spacetimes}
\lb{SecIV}
\renewcommand{\theequation}{4.\arabic{equation}}
\setcounter{equation}{0}

To study the global structure of the solutions given in the above section,
in this section we shall restrict ourselves only to the flat case, $k_{D} = 0$,
and assume that the spacetime has only four-dimensions,
$D = 4$. Then,   the surface of constant $t$ and $r$
can be a plane, a cylinder,  a M\"obius band,  a torus,
or a Klein bottle, depending on how to   identify
the two coordinates  in $X^{2}$ \cite{Wolf67}.

Setting $k = 0$ and $D =4$ in Eq.(\ref{3.16}) we find that
the corresponding metric can be written in the form,
\bq
\lb{4.1}
ds^2 = -f(x) dt^2 + f^{- 1}(x) dx^2 + x^{2} (dy^2 + dz^2),
\eq
where
\bqn
\lb{4.2}
f(x) &=& \frac{2b}{x} + \frac{q^2}{x^2} - \frac{\Lambda}{3} x^{2}, \nb\\
F_{\mu\nu} &=& \frac{q}{x^2} (\delta^{0}_{\mu}\delta^{1}_{\nu}
- \delta^{1}_{\mu}\delta^{0}_{\nu}),
\eqn
with $b \equiv - M,\; q \equiv Q$ and $\Lambda$ actually being $\Lambda_{D}$.
To consider the most general case, we assume that
the mass parameter $b$ and the coordinate $x$ take their values from the
range $- \infty < b, x < + \infty$. It is interesting to note that
if we make the replacement $(M, r) \rightarrow (-M, -r)$, we find that the
Reissner-Nordstr\"om solution coupled with a cosmological constant
remains the same. That is, the physics for $M \ge 0,\; r \ge 0$ is the same as
that for $M \le 0,\; r \le 0$. When the spacetime has plane symmetry, the range
of the coordinate $x$ is $- \infty < x < + \infty$.  This is the main
reason why here we are allowed ourselves also to consider
the case where $b > 0$.

It should be noted that the above solutions was also studied
\cite{TBHs1}. Since here we would like to give a systematical study
for this case,  some of the materials to be presented below
will be unavoidably  overlapped with some presented there.

From Eq.(\ref{4.1}) we can see that  the metric is singular on
the hypersurfaces $f(x) = 0$.
However, these singularities are coordinate singularities, except for
the one located on the hypersurface $x = 0$. This can be seen, for
example, from Eq.(\ref{3.20}) as well as the corresponding Kretschmann
scalar, which now is given by
\bq
\lb{4.3}
{\cal{R}} \equiv R^{\mu \nu \lambda \delta}
R_{\mu \nu \lambda \delta}  =
8  \left\{\frac{1}{x^{8}}\left[6\left(q^{2} + bx\right)^{2} + q^4\right]
+ \frac{1}{3} \Lambda^{2}\right\}.
\eq
The above expression shows clearly that the Kretschmann scalar becomes
unbounded only when $ x \rightarrow 0$. As $|x| \rightarrow + \infty$
the spacetime is asymptotically de Sitter or anti-de Sitter, depending
on the signs of the cosmological constant, $\Lambda$.  Thus, only the
singularity at $x = 0$ is a spacetime curvature singularity,
which divides the
whole ($t, x$)-plane into two causally disconnected regions,
$x > 0$ and $x < 0$.  All the other singularities are coordinate ones,
and to have a geodesically complete spacetime, we need to extend
the metric beyond these singularities.

In the stationary axisymmetric case, a significant quantity
that characterizes a horizon is the surface gravity \cite{NF89}.
In the following we shall derive this quantity for the spacetimes
described by the metric (\ref{4.1}) \cite{CGS93}.
 To this end, let us consider the
four-acceleration of a static observer with the four-velocity $u^{\mu}
= f^{-1/2}\delta^{\mu}_{t}$,
\bq
\lb{4.5}
a^{\mu} \equiv u^{\mu};_{\nu}u^{\nu} = \kappa(x) \delta^{\mu}_{x},
\eq
where
\bq
\lb{4.6}
\kappa(x) \equiv \frac{1}{2}f,_{x}(x).
\eq
From the above equations we can see that in order to stay at a fixed
point $x$, the observer has to fire a jet engine whose thrust per unit
mass is $|\kappa(x)|$ in the positive $x-$direction for $f,_{x}(x) >
0$, and negative $x-$direction for $f,_{x}(x) < 0$. On the horizon $x =
x_{0}$, where $f(x_{0}) = 0$, the quantity $\kappa$ defined by
\bq
\lb{4.7}
\kappa \equiv \kappa(x_{0})  = \frac{1}{2}f,_{x}\left(x_{0}\right),
\eq
is called the surface gravity of the horizon \footnote{Note the
similarity to the spherical case: $ds^{2} = f(r)dt^{2} -
f(r)^{-1}dr^{2} - r^{2}(d\theta^{2} + \sin^{2}\theta d\phi^{2})$. For a
static observer with the four-velocity $u^{\mu} =
f^{-1/2}\delta^{\mu}_{t}$, its four-acceleration is given by $a^{\mu} =
u^{\mu};_{\nu}u^{\nu} = (f,_{r}/2) \delta^{\mu}_{r}$. The quantity $\kappa
\equiv f,_{r}(r_{g})/2$ is called the surface gravity of the horizon at
$r = r_{g}$ \cite{NF89}.}.

To extend the spacetime across the horizon,  we can first introduce a new
coordinate $x^{*}$ by \cite{Walker70}
\bq
\lb{4.8}
x^{*} = \int{\frac{dx}{f(x)}},
\eq
and then two null coordinates, $v$ and $w$, via the relations
\bq
\lb{4.9}
v = e^{\kappa (t + x^{*})}, \;\;\;\;\; w = - e^{- \kappa (t - x^{*})}.
\eq
In terms of  $v$ and $w$, metric (\ref{4.1}) becomes
\bq
\lb{4.10}
ds^2 = - \kappa^{- 2}f(x) e^{- 2\kappa x^{*}}dv dw  +
x^{2} (dy^2 + dz^2).
\eq
To study the global structure of the spacetime,   it is
found convenient to distinguish the following cases:
\bqn
\lb{4.11}
A)\; &b& \not= 0, q \not= 0, \Lambda = 0; \;\;\;\;\;
B)\; b \not= 0, q = 0, \Lambda \not= 0; \nonumber\\
C)\; &b& = 0, q \not= 0, \Lambda \not= 0; \;\;\;\;\;
D)\; b \not= 0, q \not= 0, \Lambda \not= 0; \nonumber\\
E)\; &b& \not= 0, q = 0, \Lambda = 0; \;\;\;\;\;
F)\; b = 0, q \not= 0, \Lambda = 0; \nonumber\\
G)\; &b& = 0, q = 0, \Lambda \not= 0.
\eqn
In the above classification, Case E) corresponds to the Taub vacuum
solution \cite{Taub51} and the properties of which are well-known. In
particular, it has  a naked singularity at $x = 0$.  Case F) represents
a pure electromagnetic field with a naked singularity at
 $x = 0$, while Case G) corresponds to the de Sitter or
anti-de Sitter spacetime, depending on the sign of the cosmological
constant.  The global structure of the latter is studied extensively
in \cite{HE73}.  Therefore, in the following we shall focus our
attention only on the first four cases.

\subsection{ $b \not= 0, q \not= 0, \Lambda = 0.$}

When $\Lambda = 0$, we find
\bqn
\lb{4.12}
f(x) &=& \frac{2b}{x^{2}}\left(x + \frac{q^{2}}{2b}\right),\nb\\
f'(x) &=& - \frac{2b}{x^{3}}\left(x + \frac{q^{2}}{b}\right).
\eqn
Now let us consider the two cases $b < 0$ and $b > 0$ separately.
 \begin{figure}[htbp]
 \begin{center}
 \label{fig1}
 \leavevmode
  \epsfig{file=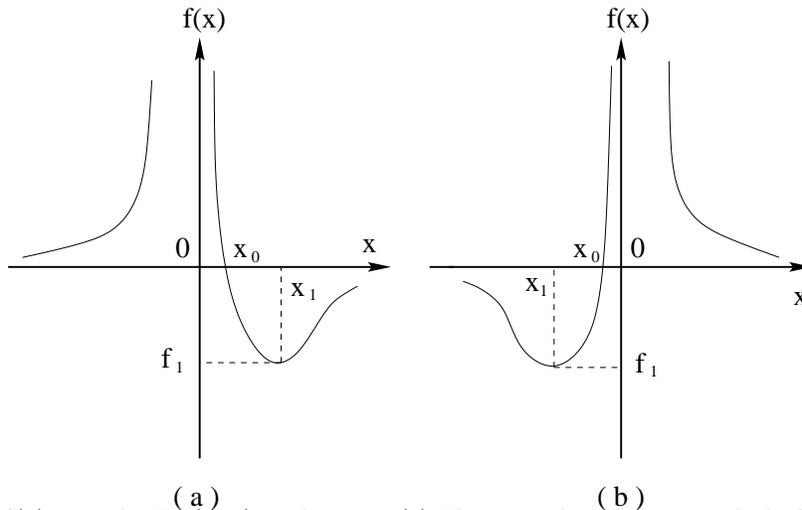,width=0.6\textwidth,angle=0}
 \caption{The function $f(x)$ given by Eq.(\ref{4.12}) in the text.
 (a) The case where $b < 0$, in which the function $f(x)$ is zero at
 $x = x_{0} \equiv q^{2}/(2|b|)$ and has a minimum at $x = x_{1}
 \equiv 2x_{0}$, where
 $f_{1}(x_{1}) = -b^{2}/q^{2}$. (b) The function $f(x)$ for $b > 0$
 and now we have $x_{0} \equiv - q^{2}/(2b) < 0$ and $f_{1}(2x_{0})
 = -b^{2}/q^{2}$.}
 \end{center}
 \end{figure}

{\bf Case A.1) $\; b < 0$}: In this case the funtion $f(x)$ is
positive in the whole region $x < 0$ [cf. Fig. 1(a)], and the
singularity at $x = 0$ is naked.
The corresponding Penrose diagram is that of Fig. 2. Note that
now $\kappa(x) = f'(x)/2 = (|bx| +q^{2})|x|^{-3}$ is also
positive in this region, so the naked singularity at $x = 0$
produces a repulsive force to a stationary
observer who stays in this region. Consequently,
the naked singularity should effectively have negative gravitational
mass.

 \begin{figure}[htbp]
 \begin{center}
 \label{fig2}
 \leavevmode
  \epsfig{file=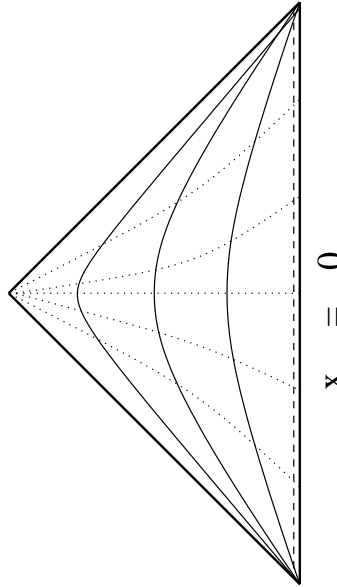,width=0.25\textwidth,angle=0}
 \caption{The Penrose diagram for the solution of Eq.(\ref{4.12}) with
 $b < 0$ in the region $x < 0$.
 The singularity at $x = 0$ is timelike and naked. The vertical curves
 are the hypersurfaces $x = Const.$ and the horizontal dotted lines are
 the ones $ t = Const.$}
 \end{center}
 \end{figure}

However, in the region $x > 0$, the function $f(x)$ becomes
zero at the point
\bq
\lb{4.13}
x_{0} =    \frac{q^{2}}{2|b|} > 0,  \;\; (b < 0),
\eq
and the corresponding surface gravity is given by
\bq
\lb{4.15}
\kappa = - \frac{4|b|^{3}}{q^{4}},\;\;\; (b < 0).
\eq
Thus,  a stationary observer near the hypersurface
$x = x_{0}$ in this region  feels  a repulsive force from the
region $0 < x < x_{0}$, and the totally effective gravitational
mass should be also negative in this region. Integrating
Eq.(\ref{4.8}) we obtain
\bq
\lb{4.14}
x^{*} =  \frac{x^2}{4b} - \frac{q^2 x}{4b^{2}} +
\frac{q^{4}}{8b^{3}}\ln\left(q^2 - 2|b|x\right).
\eq
On the other hand, the coefficient $g_{vw}$  now is given by
\bq
\lb{4.16}
g_{vw} =  - (\sqrt{2}\kappa x)^{- 2} \exp\left(- \frac{2|b| x}{q^{4}}
\left(q^{2} + |b|x\right) \right),
\eq
which  shows that the metric now is well defined
across the hypersurface $x = x_{0}$. Combining Eqs.(\ref{4.9})
and (\ref{4.14}), we also have
\bqn
\lb{4.17a}
vw &=& -\left(q^{2} - 2|b|x\right) \exp\left(\frac{2|b|^{2}}{q^{4}}x^{2}
+\frac{2|b|}{q^{2}}x \right), \\
\lb{4.17b}
v &=& \left(q^{2} - 2|b|x\right)^{1/2}\exp\left(-\frac{4|b|^{3}}{q^{4}}t +
\frac{|b|^{2}}{q^{4}}x^{2} + \frac{|b|}{q^{2}}x \right) \;\;\; (Region\;I).
\eqn
From the above expressions we can see that the coordinate
transformations (\ref{4.9}) are restricted only to the region where $v > 0$ and
$w < 0$, which will be referred to as region $I$ [cf. Fig. 3]. To extend
the metric (\ref{4.10}) to cover the whole spacetime, one simply takes the range of
$v$ and $w$ as $ - \infty < v, w < + \infty$. Then, in the $(v,
w)-$coordinates we obtain three extended regions $I', II$ and $II'$,
which are absent in the $(t, x)-$coordinates. In each of the three
extended regions, the coordinate transformations from $(v, w)$ to $(t,
x)$ are given by
\bqn
\lb{4.18}
v &=& - \left(q^{2} - 2|b|x\right)^{1/2}\exp\left(-\frac{4|b|^{3}}{q^{4}}t +
\frac{|b|^{2}}{q^{4}}x^{2} + \frac{|b|}{q^{2}}x \right), \;\;\;
(Region \;I'), \nonumber\\
v &=& \left(2|b|x - q^{2}\right)^{1/2}\exp\left(-\frac{4|b|^{3}}{q^{4}}t +
\frac{|b|^{2}}{q^{4}}x^{2} + \frac{|b|}{q^{2}}x \right), \;\;\;
(Region\; II), \nonumber\\
v &=& - \left(2|b|x - q^{2}\right)^{1/2}\exp\left(-\frac{4|b|^{3}}{q^{4}}t +
\frac{|b|^{2}}{q^{4}}x^{2} + \frac{|b|}{q^{2}}x \right),
 \;\;\; (Region\; II'),
\eqn
where $w$ can be found through Eq.(\ref{4.17a}). Then,   the corresponding Penrose
diagram is that of Fig. 3. From there we can see that the hypersurfaces $x =
x_{0}$ (or $vw = 0$) are Cauchy horizons that
separate the two regions $I$ and $I{}'$  from the two asymptotically
flat (only in the $x-$direction) ones $II$ and $II{}'$  \cite{HE73}.
The time-like coordinate $t$ is  past-directed in region $I$ and future-directed in
region $I{}'$, as we can see from Eqs.(\ref{4.17a})-(\ref{4.18}).
Across the horizons, $t$ becomes space-like, while $x$ becomes time-like.

 \begin{figure}[htbp]
 \begin{center}
 \label{fig3}
 \leavevmode
  \epsfig{file=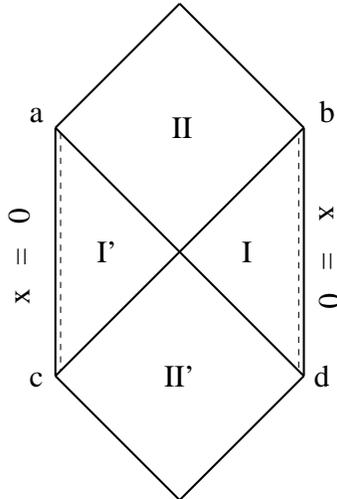,width=0.25\textwidth,angle=0}
 \caption{The Penrose diagram for the solution of Eq.(\ref{4.12}) with
 $b < 0$ in the region $x > 0$.
 The singularity at $x = 0$ is timelike and the lines $bc$ and $ad$
 where $x = x_{0}$ are Cauchy horizons. }
 \end{center}
 \end{figure}

{\bf Case A.2) $\; b > 0$}: In this case the spacetime properties can be obtained
from the case $b < 0$ by the replacement $(b, x) \rightarrow (-b, -x)$.
In particular, in the region $x > 0$ the function $f(x)$ is strictly positive
and the singularity at $x = 0$ is naked.
The corresponding Penrose diagram is similar to that given by Fig. 2,
but now with $x > 0$. In the region $x < 0$,
the metric becomes singular at the hypersurface $x = x_{0} \equiv - q^{2}/(2b)$
[cf. Fig. 1(b)], on which we now have
\bq
\lb{4.15a}
\kappa = \frac{4b^{3}}{q^{4}} > 0,\;\;\; (b >  0).
\eq
That is, a stationary observer near the hypersurface
$x = x_{0}$ in this region feels  a repulsive force from the
region $x_{0} < x < 0$. After extending the spacetime beyond this surface,
the corresponding Penrose diagram is similar to that of Fig. 3 but with $x < 0$.

\subsection{ $ b \not= 0, q = 0, \Lambda \not= 0.$}

When $q = 0$, we have
\bq
\lb{4.19}
f(x) = \frac{\Lambda}{3x}\left(\frac{6b}{\Lambda} - x^{3}\right).
\eq
In this case it is found convenient to  distinguish the four subcases,
\begin{eqnarray}
\lb{4.20}
1)\; \Lambda > 0, \;\; b > 0; \;\;\;\;
2)\; \Lambda > 0, \;\; b < 0; \nonumber\\
3)\; \Lambda < 0, \;\; b < 0; \;\;\;\;
4)\; \Lambda < 0, \;\; b > 0.
\end{eqnarray}

{\bf Case B.1) $\; \Lambda > 0, b > 0$}: In this subcase, we find that $f(x)$
is negative for all $x < 0$. Thus, in this region  $x$ is
timelike and  $t$ is spacelike. Then, the singularity at $x = 0$ is
spacelike and naked. So, in this region the spacetime has a
global structure that is quite similar to some cosmological models.

 \begin{figure}[htbp]
 \begin{center}
 \label{fig4}
 \leavevmode
  \epsfig{file=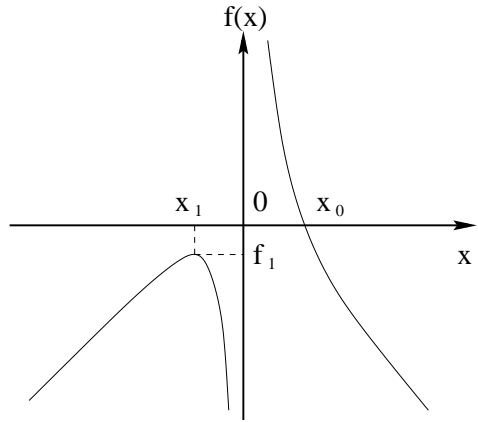,width=0.35\textwidth,angle=0}
 \caption{The function $f(x)$ given by Eq.(\ref{4.19}) in the text,
 where $x_{1} \equiv -(3b/\Lambda)^{1/3}$ and $f_{1}(x_{1})
 = - [(3b)^{2}\Lambda]^{1/3}
 < 0$. }
 \end{center}
 \end{figure}

In the region $x > 0$ the function
$f(x)$ is positive only for $ 0 \le x \le x_{0}$ [cf. Fig. 4], where
\bq
\lb{4.21}
x_{0} =  \left(\frac{6b}{\Lambda}\right)^{1/3}.
\eq
On the hypersurface $x = x_{0}$ we have $f(x_{0}) = 0$, which, as shown
before, is only a coordinate singularity. Thus, we need to extend the solution
beyond this hypersurface.  Before doing so, we first note that
now we have
\bq
\lb{4.22}
\kappa = - \frac{\Lambda}{2} x_{0} < 0,
\eq
that is, an observer near the horizon will feel a repulsive
force from the region $0< x < x_{0}$. Then, we would expect that the
singularity at $x = 0$ has an effectively  negative gravitational mass.
This is understandable, since
in the present case we have $\Lambda > 0$, which is energetically
equivalent to a matter field with a negative mass density.

Following the same line as outlined in the last subcase, we find that
the  corresponding Penrose diagram is given by Fig. 5. The hypersurfaces $x =
x_{0}$ (or $vw = 0$) are Cauchy horizons. The singularities at $x = 0$
now are also timelike.  The coordinate $t$ is
past-directed in region $I$ and future-directed in region $I'$.

 \begin{figure}[htbp]
 \begin{center}
 \label{fig5}
 \leavevmode
  \epsfig{file=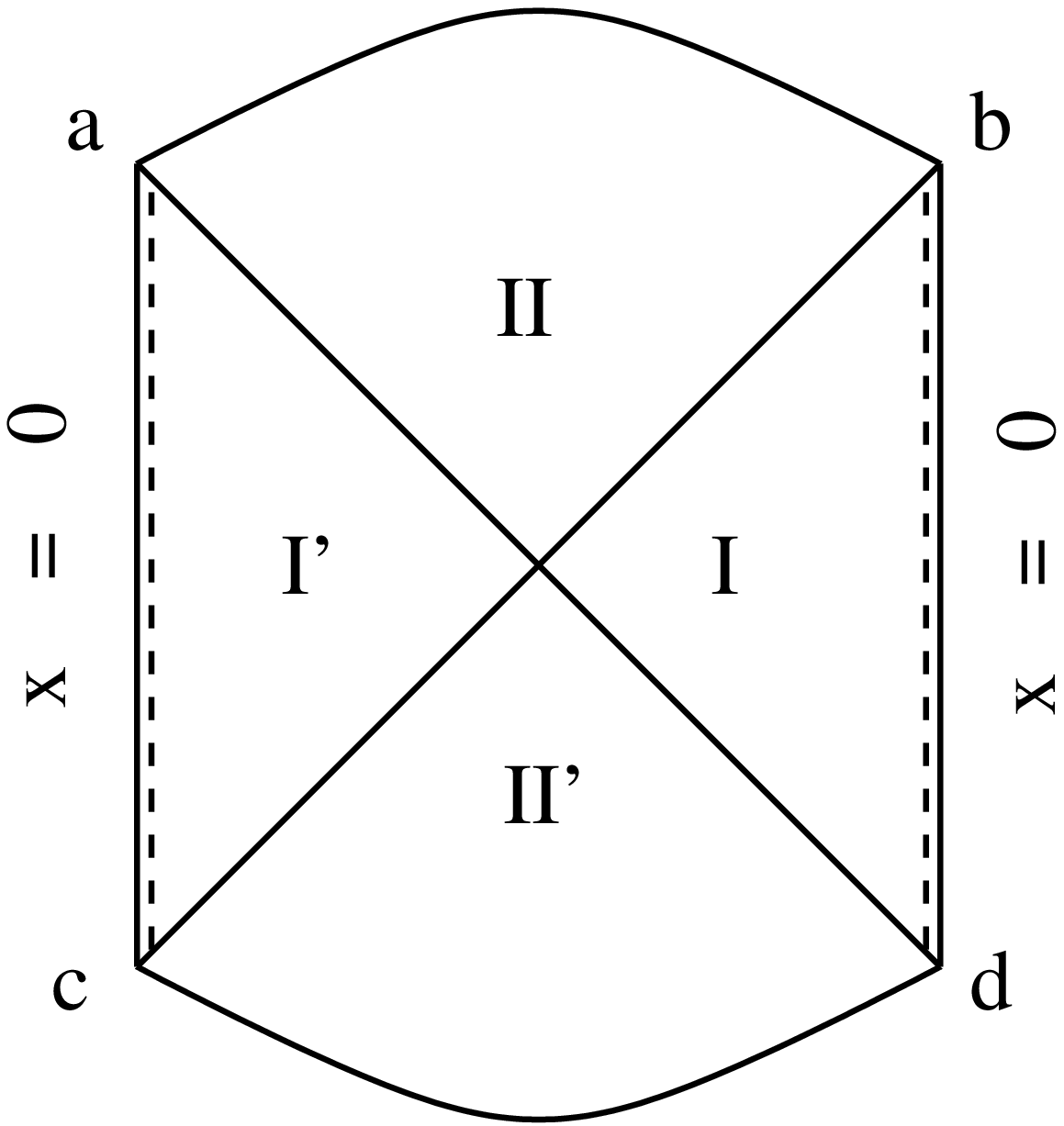,width=0.35\textwidth,angle=0}
 \caption{The Penrose diagram for the solution of Eq.(\ref{4.19}) with
 $b > 0$ and $\Lambda > 0$ in the region $x > 0$.
 The singularity at $x = 0$ is timelike and the lines $bc$ and $ad$
 where $x = x_{0}$ are Cauchy horizons. On the curved lines $ab$ and
 $cd$ we have $x = \infty$.}
 \end{center}
 \end{figure}

{\bf Case B.2) $\; \Lambda > 0, b < 0$}: From Eq.(\ref{4.19}) we can see that
this case can be obtained from the last case by the replacement
$(x, b) \rightarrow (-x, -b)$.  Thus, the spacetime structure for this case
can be also obtained from the ones given in the last case by this replacement.
In particular, now the solution in the
region $x > 0$ may represent a cosmological model, and the spacetime has
a naked singularity at $x = 0$, which is spacelike. In the region $x < 0$,
the corresponding Penrose diagram is that of Fig. 5 but now with $x < 0$.

 \begin{figure}[htbp]
 \begin{center}
 \label{fig6}
 \leavevmode
  \epsfig{file=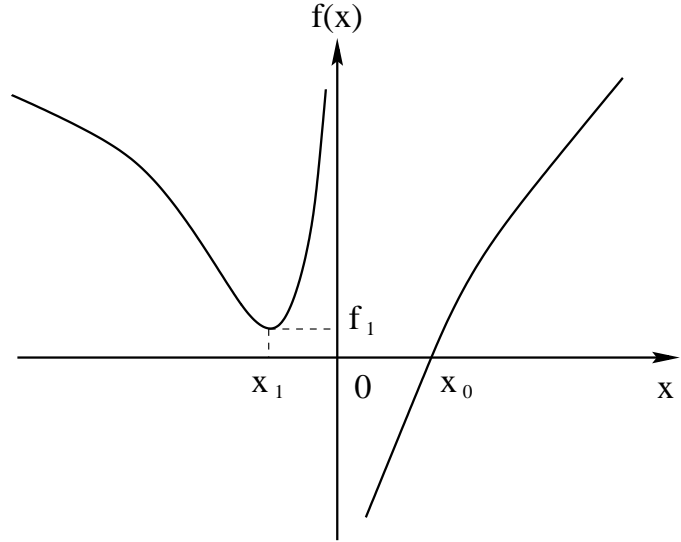,width=0.5\textwidth,angle=0}
 \caption{The function $f(x)$ given by Eq.(\ref{4.23}) in the text,
for $\Lambda < 0$ and $b < 0$, where $f_{1}(x_{1})
=  [(3b)^{2}|\Lambda|]^{1/3} >0$. }
 \end{center}
 \end{figure}
{\bf Case B.3) $\; \Lambda < 0, b < 0$}: In this case, we find that
\bqn
\lb{4.23}
f(x) &=& \frac{|\Lambda|}{3x}\left(x^{3}
- \left|\frac{6b}{\Lambda}\right|\right),\nb\\
f'(x) &=& \frac{|2\Lambda|}{3x^{2}}\left(x^{3}
+ \left|\frac{3b}{\Lambda}\right|\right).
\eqn
Thus, in the region $x < 0$ the function is always positive as shown by Fig. 6.
As a result, the singularity at $x = 0$ is naked. On the other hand, for any
hypersurface $x = C$, its normal vector $n_{\alpha}$ is given by
$n_{\alpha} = \partial(x - C)/\partial x^{\alpha} = \delta^{x}_{\mu}$. Thus, we
have
\bq
\lb{4.23a}
x_{\alpha} x^{\alpha} = f(x) \rightarrow + \infty,
\eq
as $ x \rightarrow - \infty$, which means that the spatial infinity $x = - \infty$
in the present case is timelike. Then, the corresponding Penrose diagram is given by
Fig. 7.

It is also interesting  to note that now the acceleration of a static observer
is given by Eq.(\ref{4.5}) with
\bq
\lb{4.24}
\kappa(x) = \frac{|\Lambda|}{3x^{2}}\left(x^{3}
+ \left|\frac{3b}{\Lambda}\right|\right)
= \cases{ < 0, & $ x < x_{1}$,\cr
= 0, & $ x = x_{1}$,\cr
> 0, & $ x_{1} < x < 0$,\cr}
\eq
where $x_{1} \equiv - |3b/\Lambda|^{1/3}$.

 \begin{figure}[htbp]
 \begin{center}
 \label{fig6a}
 \leavevmode
\epsfig{file=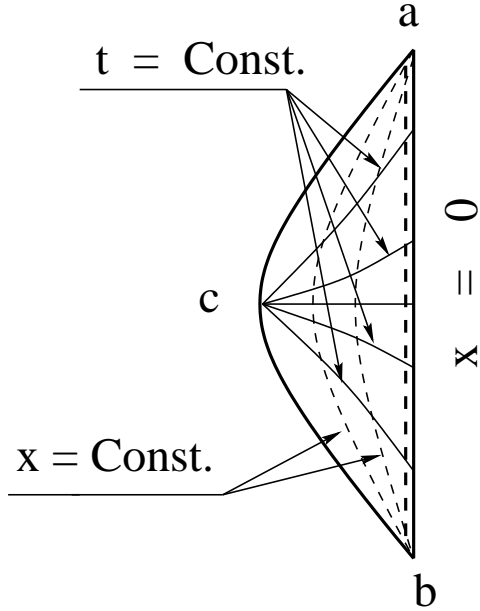,width=0.35\textwidth,angle=0}
\caption{The Penrose diagram for the solutions given by Eq.(\ref{4.23})
 in the region $ x < 0$ for $\Lambda < 0$ and $b < 0$. }
 \end{center}
 \end{figure}

In the region $x > 0$, we have
\bq
\lb{4.25}
f(x) =  \cases{ < 0, & $ x < x_{0}$,\cr
= 0, & $ x = x_{0}$,\cr
> 0, & $ x > x_{0}$,\cr}
\eq
where $x_{0} \equiv |6b/\Lambda|^{1/3}$, and
\bq
\lb{4.26}
\kappa \equiv \frac{1}{2}f'(x_{0}) = \left|\frac{\Lambda}{2}\right|x_{0} > 0.
\eq

Extending the solutions into the region $0 < x < x_{0}$, we find the following
relevant quantities
\begin{eqnarray}
\lb{4.27}
g_{vw} &=& - \frac{2(x^2 + x_{0}x + x_{0}^2)^{3/2}}
{3|\Lambda| x_{0}^2 x}
\exp\left\{- \sqrt{3}{\; \mbox{arctg}}\left(\frac{2x +
x_{0}}{\sqrt{3} x_{0}}\right)\right\}, \nonumber\\
vw &=& -\frac{x - x_{0}}{ (x^2 + x_{0}x + x_{0}^2)^{1/2}}
\exp\left\{\sqrt{3}{\; \mbox{ arctg}}\left(\frac{2x +
x_{0}}{\sqrt{3} x_{0}}\right)\right\},\\
v &=& \frac{(x - x_{0})^{1/2}}{ (x^2 + x_{0}x +
x_{0}^2)^{1/4}}\nonumber\\
& & \exp\left\{ - \frac{\Lambda}{2}x_{0}t +
\frac{\sqrt{3}}{2}{\; \mbox{ arctg}}\left(\frac{2x +
x_{0}}{\sqrt{3} x_{0}}\right)\right\} \; (Region\;I).
\end{eqnarray}
The corresponding Penrose diagram is given by Fig. 8, which shows that now
the hypersurfaces $x = x_{0}$ (or $vw = 0$) represent event horizons, and the
singularities at $x = 0$ are space-like. Regions $II$ and $II'$ are
two catastrophic regions, while regions $I$ and $I'$ are  two
asymptotically anti-de Sitter regions. The time-like coordinates $t$ is
future-directed in region $I$ and past-directed in region $I'$.  Region
$II$ can be considered as a black hole, while region $II'$ as a white
hole.  Since now the surface gravity of the
horizon is positive, the spacetime singularity at $x = 0$ is
gravitationally attractive and expected to have an effectively positive
gravitational mass.

 \begin{figure}[htbp]
 \begin{center}
 \label{fig7}
 \leavevmode
  \epsfig{file=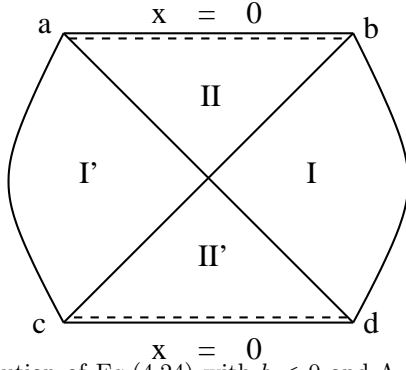,width=0.3\textwidth,angle=0}
 \caption{The Penrose diagram for the solution of Eq.(\ref{4.23}) with
 $b < 0$ and $\Lambda < 0$ in the region $x > 0$.
 The singularity at $x = 0$ is spacelike and the lines $bc$ and $ad$
 where $x = x_{0}$ represent event horizons, while the curved lines
 $ac$ and $bd$ represent the spatial infinity $ x = \infty$.}
 \end{center}
 \end{figure}

{\bf Case B.4) $\; \Lambda < 0, b > 0$}: This is the case which corresponds to Case
B.3).  As a matter of fact, replacing $x$ by $- x$, we can obtain one
from the other. Therefore, the corresponding Penrose diagrams in the two subcases
are the same, after the above replacement is taken into account.

\subsection{ $ b = 0, q \not= 0, \Lambda \not= 0.$}

When $b = 0$, from Eq.(\ref{4.2}) we can see that $f(x) = f(-x)$. Thus,
the spacetime has reflection symmetry with respect to the hypersurface
$x = 0$. Then, without loss of generality, in the following we shall
only consider the region $x > 0$ in this case.

When  $\Lambda < 0$, $f(x)$ is always greater than zero for any $x$. Consequently,
the singularity at $x = 0$ is naked, and the corresponding Penrose diagram is
similar to that of Fig. 7.

When $\Lambda > 0$, we have
\bq
\lb{4.28}
f(x) = \frac{\Lambda}{3x^{2}} (x_{0}^2 + x^{2})(x_{0}^2 - x^{2}),
\eq
but now $x_{0}$ is given by
\bq
\lb{4.29}
x_{0} \equiv \left(\frac{3q^{2}}{\Lambda}\right)^{1/4}.
\eq
From the above equations we can see that $f(x) \ge 0$ holds only in the
region $0 < x < x_{0}$. On the   hypersurface  $x =  x_{0}$ the surface
gravity is given by
\bq
\lb{4.30}
\kappa = - \frac{2\Lambda}{3} x_{0} < 0,
\eq
that is, the effective gravitational mass in the region $0 < x < x_{0}$
should be negative. One can show that the corresponding Penrose diagram
is similar to that of Fig. 5.

\subsection{ $ b \not= 0, q \not= 0, \Lambda \not= 0.$}

When none of the three parameters are zero, we need to distinguish the
four subcases as defined by Eq.(\ref{4.20}). However, as shown in Cases A) and
B), the properties of the spacetimes for the case $b < 0$ can be
obtained from those for the case $b > 0$ by simply replacing $x$ by
$- x$. Thus, we actually need to consider only the subcases $\Lambda > 0, \; b >
0$, and  $\Lambda < 0, \; b < 0$.

{\bf Case D.1) $\; \Lambda > 0, \; b > 0$}: In this case, it can be shown that
$f(x)$ takes the form
\bq
\lb{4.31}
f(x) = \frac{\Lambda}{3x^{2}}(x_{+} - x)(x - x_{-})(x^{2} + Bx + C),
\eq
where
\bq
\lb{4.32}
B \equiv x_{+} + x_{-},\;\;\;\;
 C \equiv x_{+}^{2} + x_{+}x_{-} + x_{-}^{2},
 \eq
and $x_{\pm}$ are the two  real roots of the equation $f(x) = 0$, given
via the relations
\bq
\lb{4.33}
B (x_{+}^{2} + x_{-}^{2})  = \frac{6b}{\Lambda}, \;\;\;
 C x_{+}x_{-}  = - \frac{3q^{2}}{\Lambda},
\eq
with the property
\bq
\lb{4.34}
x_{+} > 0, \;\;\;\;\;\;\;\;\;  x_{-}< 0.
 \eq
From the above equations we can see that $f(x) \ge 0$ is true only in
the region $ x_{-} \le x < 0$ or $0 < x \le x_{+}$. Then, we
need to extend the solutions beyond both of the hypersurfaces $x = x_{\pm}$.

Let us first consider the extension  across the hypersurface $x = x_{+}$.
We first note that
\bq
\lb{4.35a}
\left. \kappa\right|_{x = x_{+}}  =
- \frac{\Lambda}{6x_{+}^{2}}(x_{+} - x_{-})
(x_{+}^{2} + Bx_{+} + C) < 0.
\eq
Thus, a static observer near the horizon will feel a repulsive force
from the region $ 0 < x < x_{+}$, which means that the effective
gravitational mass in this region is negative. One can also show that
the Penrose diagram is similar to that given by Fig. 5.

On the other hand, the surface gravity at $x = x_{-}$ is given by
\bq
\lb{4.35b}
\left. \kappa\right|_{x = x_{-}} =
\frac{\Lambda}{6x_{-}^{2}}(x_{+} - x_{-}) (x_{-}^{2} + Bx_{-} + C) > 0,
\eq
which means, similar to that in the region $x > 0$, now a static
observer near the horizon will also feel a repulsive force
from the region $ x_{-} < x < 0$. It can be also shown that the
corresponding Penrose diagram is that of Fig. 5,  but with
$x < 0$.

{\bf Case D.2) $\; \Lambda < 0,\; b < 0$}: In this case we find that it
is convenient to
distinguish the following three subcases:
\bq
\lb{4.36}
i)\; |q| < q_{c}, \;\;\;\;\;\;
ii)\; |q| = q_{c}, \;\;\;\;\;\;
iii)\; |q| > q_{c},
\eq
where
\bq
\lb{4.37}
 q_{c} \equiv |\Lambda|^{-1/6}\left|\frac{3b}{2}\right|^{2/3}.
\eq

{\bf Case D.2.i) $\; |q| < q_{c}$}: In this case, it is found that the
function $f(x)$ can be written as
\bq
\lb{4.38}
f(x) = \frac{|\Lambda|}{3x^{2}}(x - x_{+})(x - x_{-})(x^{2} + Bx + C),
\eq
where $B, C$ and $x_{\pm}$  are  given by Eqs.(\ref{4.32}) and (\ref{4.33}),
but now with the properties
 \bq
 \lb{4.39}
x_{+} > 0, \;\;\; x_{-} > 0, \;\; (|q| < q_{c}).
\eq
From Eqs.(\ref{4.38}) and (\ref{4.39}) we can see that in the region
$x < 0$ the function $f(x)$ is positive, and the singularity at $x = 0$
is naked  [cf. Fig. 9]. The corresponding Penrose diagram is similar
to that given  by  Fig. 7.

 \begin{figure}[htbp]
 \begin{center}
 \label{fig8}
 \leavevmode
  \epsfig{file=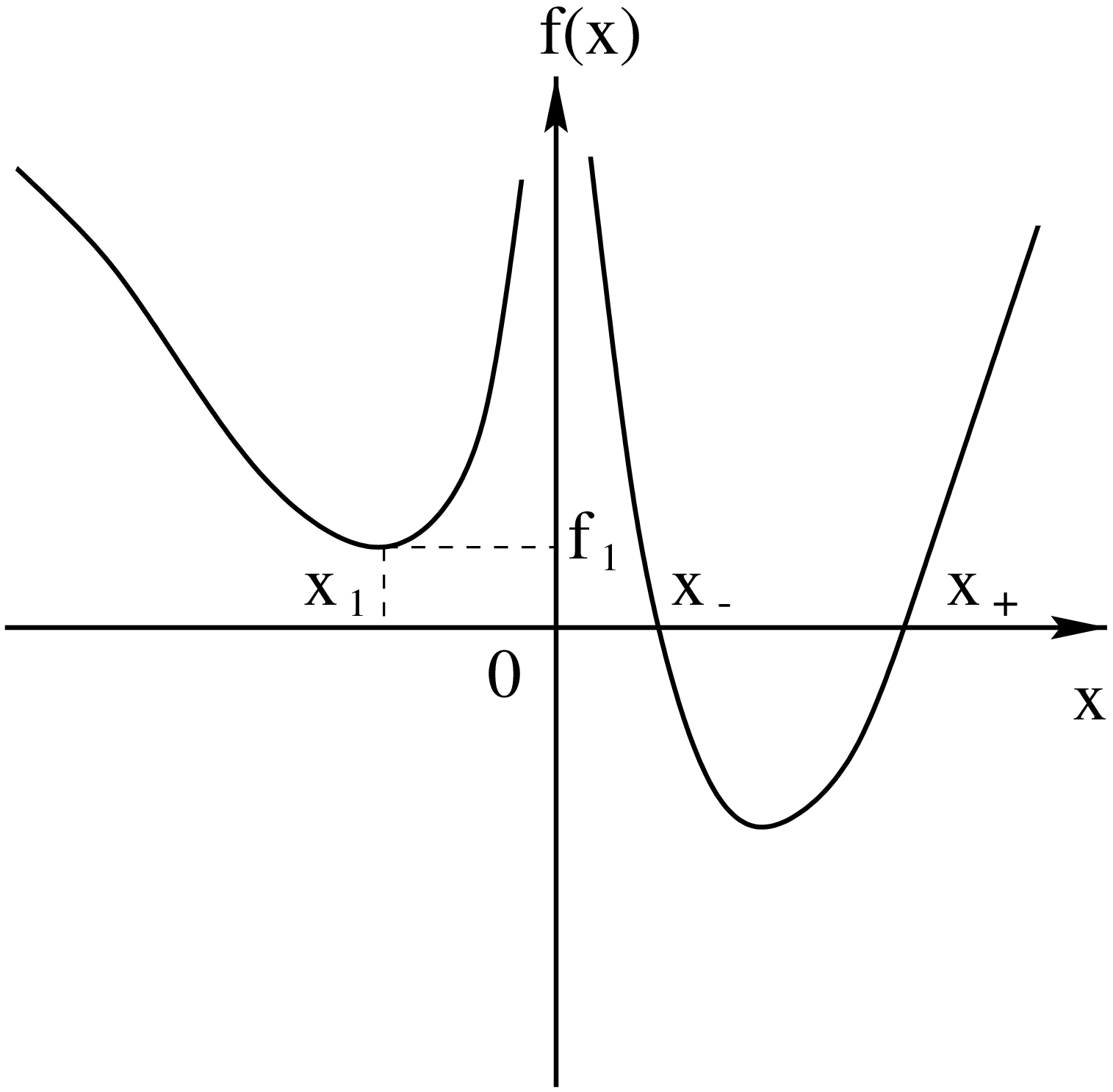,width=0.35\textwidth,angle=0}
 \caption{The function $f(x)$ given by Eq.(\ref{4.38}) in the text,
for $\Lambda < 0,\; b < 0$ and $|q| < q_{c}$, where $f_{1}(x_{1})
=  [(3b)^{2}|\Lambda|]^{1/3} >0$. }
 \end{center}
 \end{figure}

When $ x > 0$, $f(x)$ is non-negative  only in the regions $ x \ge x_{+}$ or
$0 \le x \le x_{-}$.  Thus, an extension across the hypersurfaces $x = x_{\pm}$
is needed. Let us first consider the extension across $x = x_{+}$.

{\bf The extension across the hypersurface} $x = x_{+}$: Following previous
cases,   we find the following,
\begin{eqnarray}
\lb{4.40}
x^{*} &=&  \frac{1}{2\kappa_{+}}\ln(x - x_{+}) + \sum^{\infty}_{n = 1}
{\frac{a_{n}}{n}(x - x_{+})^{n}},  \nonumber\\
\kappa_{+} &=& \frac{f,_{x}(x_{+})}{2} =
\frac{|\Lambda|}{6x_{+}^{2}}(x_{+} - x_{-})
(x_{+}^{2} + bx_{+} + c) > 0, \nonumber\\
f(x) &=& \frac{x - x_{+}}{g(x)}, \;\;\;\;\;
a_{n} = \left.\frac{1}{n!}\frac{d^{n}g(x)}{dx^{n}}\right|_{x = x_{+}},\nonumber\\
g_{vw} &=& - [2\kappa_{+}^{2}g(x)]^{-1}\exp\left\{- 2\kappa_{+}
\sum^{\infty}_{n = 1} {\frac{a_{n}}{n}(x - x_{+})^{n}}\right\}, \nonumber\\
vw &=& - (x - x_{+}) \exp\left\{2\kappa_{+}
\sum^{\infty}_{n = 1} {\frac{a_{n}}{n}(x - x_{+})^{n}}\right\}, \nonumber\\
v &=& (x - x_{+})^{1/2}
 \exp\left\{\kappa_{+}\left[t + \sum^{\infty}_{n = 1}
{\frac{a_{n}}{n}(x - x_{+})^{n}}\right] \right\}, (Region \; I).
\end{eqnarray}
Then, we get four regions, $I, I', II$ and $II'$ in the ($v, w$)-plane, as shown
in Fig. 10. The hypersurfaces $x = x_{+}$ now are event horizons. The
coordinate $t$ is future-directed in region $I$ and past-directed in
region $I'$. Note that as $x \rightarrow + \infty$, we have $f(x)
\rightarrow + \infty$. Therefore, the hypersurface $x = + \infty$ is
timelike and   represented by the curves $ac$ and $bd$   in   Fig. 10.

{\bf The extension across the hypersurface} $x = x_{-}$: Now we
find that
\begin{eqnarray}
\lb{4.41}
x^{*} &=&  \frac{1}{2\kappa_{-}}\ln(x - x_{-}) + \sum^{\infty}_{n = 1}
{\frac{b_{n}}{n}(x - x_{-})^{n}},  \nonumber\\
\kappa_{-} &=& \frac{f,_{x}(x_{-})}{2} = -
\frac{|\Lambda|}{6x_{-}^{2}}(x_{+} - x_{-})
(x_{-}^{2} + bx_{-} + c) < 0, \nonumber\\
f(x) &=& \frac{x - x_{-}}{\bar{g}(x)},\;\;\;\;\;\;
b_{n} = \left. \frac{1}{n!}\frac{d^{n}\bar{g}(x)}{dx^{n}}\right|_{x =x_{-}},\nb\\
g_{vw} &=&  - [2\kappa_{-}^{2}\bar{g}(x)]^{-1}\exp\left\{ - 2\kappa_{-}
\sum^{\infty}_{n = 1} {\frac{b_{n}}{n}(x - x_{-})^{n}}\right\}, \nonumber\\
vw &=& - (x - x_{-}) \exp\left\{- 2\kappa_{-}
\sum^{\infty}_{n = 1} {\frac{b_{n}}{n}(x - x_{-})^{n}}\right\}, \nonumber\\
v &=& (x - x_{-})^{1/2}
 \exp\left\{\kappa_{-}\left[t + \sum^{\infty}_{n = 1}
{\frac{b_{n}}{n}(x - x_{-})^{n}}\right] \right\}, (Region \; III).
\end{eqnarray}
Clearly, by this extension, we get two more regions $III$ and $III'$, as shown in
Fig. 10. The singularities at $x = 0$ are time-like and repulsive, since
now we have $\kappa_{-} < 0$. The hypersurfaces $x = x_{-}$ represent
Cauchy horizons. Extensions across these hypersurfaces are further
needed in order to have a geodesically complete spacetime.  However,
since these surfaces are Cauchy horizons, the extensions across them
are not unique, a situation quite similar to that  of
the Reissner-Nordstrom (RN) or Kerr solution \cite{HE73}.

 \begin{figure}[htbp]
 \begin{center}
 \label{fig9}
 \leavevmode
  \epsfig{file=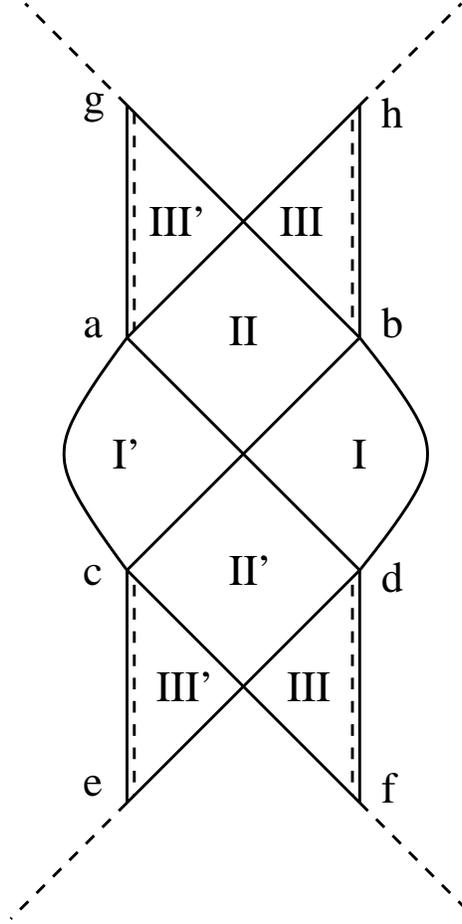,width=0.35\textwidth,angle=0}
 \caption{The Penrose diagram for the solution of Eq.(\ref{4.38}) with
 $|q| < q_{c}$ in the region $x > 0$. The lines $ad$ and $bc$
 where $x = x_{+}$ represent event horizons, and the ones
 $gb,\; ha,\; cf$ and $de$, where $x = x_{-}$, represent Cauchy
 horizons. The singularities at $hb,\; ga,\; df$ and $ce$ are timelike.
 The curves $ac$ and $bd$ represent the spatial infinity ($x = \infty$),
 which is timelike. To have a geodesically maximal spacetime, the extension
 in the vertical
 directions should be continuous to infinity.}
 \end{center}
 \end{figure}

{\bf Case D.2.ii) $\; |q| = q_{c}$}: In this case, we find
\bq
\lb{4.42}
f(x) = \frac{|\Lambda|}{3x^{2}}(x - x_{0})^{2}(x^{2} + 2 x_{0}x +
3 x_{0}^{2}),
\eq
where $x_{0}$ now is given by
\bq
\lb{4.43}
x_{0} = \left(\frac{3b}{2\Lambda}\right)^{1/3}.
\eq
Thus, $f(x) > 0$ holds for all $x < 0$. As a result, in the region $x < 0$
the singularity at $x = 0$ is naked, and the corresponding Penrose diagram
is that of Fig. 7.

In the region $x > 0$, $f(x)$ is also positive, except at the point $x = x_{0}$,
where $f(x_{0}) = 0$. See Fig. 9, and notice that now we have
$x_{+} = x_{-} = x_{0}$. This is quite similar to the extreme case of the RN
solution \cite{HE73}. Integrating Eq.(\ref{4.8}), we obtain
\bqn
\lb{4.44}
x^{*} &=&  - \frac{x}{2|\Lambda|x_{0}(x - x_{0})} +
\frac{1}{3|\Lambda|x_{0}} \ln\frac{(x - x_{0})^{2}}{x^{2} + 2 x_{0}x + 3
x_{0}^{2}}\nb\\
& &
+ \frac{7\sqrt{2}}{12 |\Lambda| x_{0}} {\; \mbox{ arctg}}\left(\frac{x +
x_{0}}{\sqrt{2}x_{0}}\right)  + x^{*}_{0},
\eqn
where $x^{*}_{0}$ is an integration constant. In the following,
without loss of generality, we shall choose $x^{*}_{0}$ such that $x^{*}(0)
= 0$. On the other hand, from Eqs.(\ref{4.7}) and (\ref{4.42}) we find
\bq
\lb{4.45}
\kappa = \frac{1}{2}f,_{x}(x_{0}) = 0.
\eq
That is, the surface gravity now is zero. As a result, the usual method
of the extension given by Eqs.(\ref{4.8})-(\ref{4.10}) is
not applicable to this case,
and we have to consider other possibilities. Following Carter
\cite{Carter66}, let us define the two null coordinates $v $ and $w$ by
\bq
\lb{4.46}
v = tan^{-1}( t + x^{*}), \;\;\;\;\; w = tan^{-1}( t - x^{*}),
\eq
where $- \pi/2 \le v \le 3\pi/2$, and $- 3\pi/2 \le w \le \pi/2$.
In terms of $v$ and $w$, the metric reads
\bq
\lb{4.47}
ds^{2} = - \frac{f(x)}{\cos^{2}v \cos^{2}w} dv dw + x^{2}( dy^{2} + dz^{2}).
\eq
Using the relation
\bq
\frac{\sin^{2}(v - w)}{\cos^{2}v \cos^{2}w} = 4(x^{*})^{2},
\eq
one can show that the metric coefficients of Eq.(\ref{4.47}) become regular
across the hypersurface $x = x_{0}$. Thus, it represents an extension.
As a matter of fact, one can show that this extension is analytic and
the extended spacetime is given by regions $I$ and $III$ in Fig. 11,
which is quite similar to the extreme RN black hole
\cite{HE73}.  The only difference is that in the present case
  the hypersurface
$ x = + \infty$ is time-like, since as $x \rightarrow + \infty$ we
now have $x^{*} \rightarrow $ finite. Keeping this in mind, we can deduce
all the properties of this solution from the extreme RN black hole. For
example, the horizons $x  = x_{0}$ represent event horizons. The region
$ x_{0} < x < + \infty$ is mapped to region $I$, while the one $ 0 < x
< x_{0}$ to region $III$. The $t$-coordinate is future-directed in
  region $I$. The singularity at $ x = 0$, represented by the vertical
double-lines in Fig. 11, is time-like. From the diagram we can see that
the extension should be continuous to infinity in the vertical
directions, in order to get a geodesically complete spacetime.

 \begin{figure}[htbp]
 \begin{center}
 \label{fig10}
 \leavevmode
  \epsfig{file=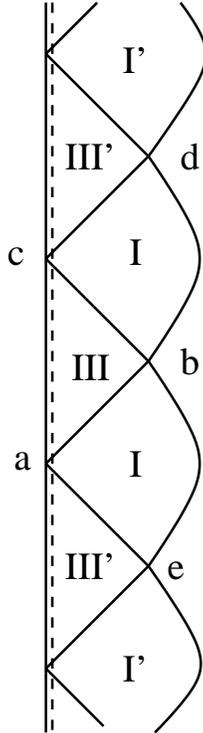,width=0.15\textwidth,angle=0}
\caption{The Penrose diagram for the solutions given by
Eq.(\ref{4.42}) in the text with   $|q| = q_{c}$ and in the
region $x > 0$. The spacetime singularity at $x = 0$,
represented by the vertical
line $ac$,  is timelike. The surfaces $ae, ab, cb,$ and $cd$,
where $x  = x_{0}$, all represent event horizons. The hypersurfaces $x
= + \infty$ now are timelike, given by the curves $bd$ and $eb$. The
extension in the vertical directions is continuous to infinity. }
 \end{center}
 \end{figure}

{\bf Case D.2.iii) $\;|q| > q_{c}$}: In this case it can be shown that
$f(x) > 0$ holds for any $x$. As a result, the singularity at $x = 0$ is naked,
and the corresponding Penrose diagram is given by Fig. 7.

\section{Type II Fluid in D-Dimensional Spacetimes}
\lb{SecV}
\renewcommand{\theequation}{5.\arabic{equation}}
\setcounter{equation}{0}

Now using the gauge freedom of Eq.(\ref{2.1a}), we shall set
\bq
\lb{5.1}
g_{11}(x^{c}) = 0,\;\;\; s = r,
\eq
where $x^{a} = \{v, r\}$. These coordinates are usually called
{\em Eddington-Finkelstein coordinates},   in terms of which   the metric
(\ref{2.1}) takes the form,
\bq
\lb{5.2}
ds^{2} = -e^{\psi(v,r)}dv\left(f(v,r)e^{\psi(v,r)}dv + 2\epsilon dr\right)
+ r^{2}h_{AB}\left(x^{C}\right)dx^{A}dx^{B},
\eq
where $\epsilon = \pm 1$. When $\epsilon = 1$, the radial coordinate $r$
increases towards the future along a ray $v =  Const$.
 When $\epsilon = -1$, the radial coordinate $r$
decreases towards the future along a ray $v =  Const$.

From Eq.(\ref{2.9}) it can be shown that
\bqn
\lb{5.4}
{}^{(2)}R_{ab} &=& \frac{1}{2}A\left[fe^{\psi}\delta^{0}_{a}\delta^{0}_{b}
+ \epsilon\left(\delta^{0}_{a}\delta^{1}_{b}
+ \delta^{1}_{a}\delta^{0}_{b}\right)\right],\nb\\
A(v,r) &=&e^{\psi}\left[2f\psi_{,rr} + f_{,rr}
+ \left(2f\psi_{,r} + 3f_{,r}\right)\psi_{,r}\right]
- 2\epsilon\psi_{,vr},\nb\\
\nabla_{a}\nabla_{b}r &=& - \frac{1}{2}fe^{2\psi}\left(2f\psi_{,r}
+ f_{,r} + \epsilon e^{-\psi}\frac{f_{,v}}{f}\right)
\delta^{0}_{a}\delta^{0}_{b}\nb\\
& & - \frac{1}{2}\epsilon e^{\psi}\left(2f\psi_{,r}
+ f_{,r}\right)\left(\delta^{0}_{a}\delta^{1}_{b}
+ \delta^{1}_{a}\delta^{0}_{b}\right)
 -\psi_{,r}\delta^{1}_{a}\delta^{1}_{b},\nb\\
\Box r &=& f\psi_{,r} + f_{,r}.
\eqn
Inserting the above expressions into Eq.(\ref{2.7}), we obtain
\bqn
\lb{5.6a}
{}^{(D)}R_{vv} &=& \frac{1}{2}f e^{2\psi}\left[2f\psi_{,rr}
+ f_{,rr} + \left(2f\psi_{,r} + 3f_{,r}\right)\psi_{,r}
+ \frac{D-2}{r}\left(2f\psi_{,r} + f_{,r}\right)\right]\nb\\
& & - \frac{1}{2}\epsilon f e^{\psi}\left(2\psi_{,rv}
-\frac{(D-2)f_{,v}}{rf}\right),\\
\lb{5.6b}
{}^{(D)}R_{vr} &=& \frac{1}{2}\epsilon e^{\psi}\left[2f\psi_{,rr}
+ f_{,rr} + \left(2f\psi_{,r} + 3f_{,r}\right)\psi_{,r}
+ \frac{D-2}{r}\left(2f\psi_{,r} + f_{,r}\right)\right]
-\psi_{,rv},  \\
\lb{5.6c}
{}^{(D)}R_{rr} &=&  \frac{D-2}{r}\psi_{,r},\\
\lb{5.6d}
{}^{(D)}R_{AB} &=& \left[k_{D} - r\left(f\psi_{,r} + f_{,r}\right)
- (D-3)f\right]h_{AB}.
\eqn

On the other hand, introducing two null vectors $l_{\mu}$ and $n_{\mu}$
via the relations
\bqn
\lb{5.7}
l_{\mu} &=& \delta^{v}_{\mu},\;\;\;\;
n_{\mu} = e^{\psi}\left(\frac{1}{2}fe^{\psi}\delta^{v}_{\mu}
+ \epsilon \delta^{r}_{\mu}\right),\nb\\
l_{\lambda}l^{\lambda} &=& 0 = n_{\lambda}n^{\lambda},\;\;\;\;
l_{\lambda}n^{\lambda} = -1,
\eqn
we find that the energy-momentum tensor for a Type $II$ fluid can be
written as  \cite{Hus96,WW99}
\bq
\lb{5.8}
{}^{(D)}T_{\mu\nu} = \mu {l}_{\mu}l_{\nu}
+ (\rho + P)\left({l}_{\mu}n_{\nu}  + {l}_{\nu}n_{\mu}\right)
+ P g_{\mu\nu}.
\eq
Then, from the Einstein field equations  we find that
\bq
\lb{5.9}
{}^{(D)}R_{\mu\nu} =   8\pi G_{D}\left\{\mu {l}_{\mu}l_{\nu}
+ (\rho + P)\left({l}_{\mu}n_{\nu}  + {l}_{\nu}n_{\mu}\right)
+ \frac{2\rho}{D-2} g_{\mu\nu}\right\}.
\eq
From Eqs.(\ref{5.6c}) and (\ref{5.9}) we obtain $\psi_{,r} = 0$,
that is, $\psi(v, r) = \psi(v)$. However, by introducing another null
coordinate $\bar{v} = \int{e^{\psi(v)}dv}$,  we can always set, without
loss of the generality,
\bq
\lb{5.10}
\psi(v, r) = 0, \;\;\; \left({}^{(D)}G_{rr} = 8\pi G_{D}{}^{(D)}T_{rr}\right).
\eq
On the other hand, from the rest of the Einstein field equations, we find that
\bqn
\lb{5.11}
\mu &=& \frac{\epsilon\left(D-2\right)}{16\pi G_{D} r} f_{,v},\nb\\
\rho &=& \frac{D-2}{16\pi G_{D} r^{2}} \left(k_{D} - rf_{,r}
- \left(D-3\right)f\right),  \nb\\
P &=& \frac{1}{16\pi G_{D}} \left\{f_{,rr} + \frac{2(D-3)}{r}f_{,r}
- \frac{D-4}{r}\left[k_{D}  - \left(D-3\right)f\right]\right\}.
\eqn

To see that the energy-momentum tensor given by Eq.(\ref{5.8}) indeed
represents a Type $II$  fluid, let us first introduce the following
unit vectors,
\bqn
\lb{5.12}
E_{(0)}^{\mu} &=& \frac{l^{\mu} + n^{\mu}}{\sqrt{2}},\;\;\;
E_{(1)}^{\mu} = \frac{l^{\mu} - n^{\mu}}{\sqrt{2}},\nb\\
E_{(A)}^{\mu} &=&\frac{1}{r}E_{(A)}^{B}\delta^{\mu}_{B},\;\;\;
g_{\mu\nu} E_{(\alpha)}^{\mu} E_{(\beta)}^{\mu} = \eta_{\alpha\beta},
\eqn
where $\eta_{\alpha\beta} = {\mbox{diag.}}\{-1, 1, 1, ..., 1\}$, and
$E_{(A)}^{B}\; (A, B = 2, 3, ..., D-2)$ consist
of an orthogonal base in the space of $h_{AB}$, that is,
\bq
\lb{5.13}
E_{(A)}^{C} E_{(B)}^{D}h_{CD} = \delta_{AB}.
\eq
Then, projecting $T_{\mu\nu}$ into this base, we find that
\bq
\lb{5.14}
\left(T_{(\mu)(\nu)}\right) =
\left(\matrix{\frac{1}{2}(\mu + 2\rho) & - \frac{1}{2}\mu & 0 & 0 & ... &0\cr
-\frac{1}{2}\mu  &  \frac{1}{2}(\mu - 2\rho) & 0 & 0 & ... &0\cr
0 & 0 & P & 0 & ... &0\cr
0 & 0 & 0 & P & ... &0\cr
... & ... & ... & ... & ... &0\cr
0 & 0 & 0 & 0 & ... &P\cr}\right),
\eq
which is exactly the Type $II$ fluid, according to the classification given in
\cite{HE73}. Therefore, {\em for any given function $f(v,r)$ with $\psi = 0$, the
corresponding energy-momentum tensor of the metric (\ref{5.2})
can be always written in the form of a Type $II$  fluid}. However, not for
any choice of $f(v,r)$ the solution is physically acceptable. In fact,  it must
satisfy some (geometrical and physical)  conditions, such as, the energy
conditions \cite{HE73}. Following \cite{WW99}, we write $f(v,r)$ in the form
\bqn
\lb{5.15}
f(v,r) &=& k - \frac{2m(v,r)}{r^{D-3}},\nb\\
m(v,r) &=& \sum^{\infty}_{n = -\infty}{a_{n}(v) r^{n}}.
\eqn
Then, Eq.(\ref{5.11}) becomes
\bqn
\lb{5.16}
\mu &=& - \frac{\epsilon\left(D-2\right)}{8\pi G_{D} r^{D-2}}
\sum^{\infty}_{n = -\infty}{{a'}_{n}(v) r^{n}},\nb\\
\rho &=& \frac{D-2}{8\pi G_{D} r^{D-2}}
\sum^{\infty}_{n = -\infty}{na_{n}(v) r^{n-1}},\nb\\
P &=& - \frac{1}{8\pi G_{D} r^{D-3}}
\sum^{\infty}_{n = -\infty}{n(n-1)a_{n}(v) r^{n-2}},
\eqn
where a prime denotes the ordinary differentiation with respect to the indicated
argument. Clearly, by properly choosing the coefficients $a_{n}(v)$, we can get
many exact solutions. For example, for the choice,
\bq
\lb{5.17}
a_{n}(v) = \cases{m(v), & $ n = 0$, \cr
-\frac{1}{2}q^{2}(v), &$ n = -(D-3)$,\cr
\frac{1}{2}\Lambda, & $ n = D-1$,\cr
0, & otherwise,\cr}
\eq
we find that
\bq
\lb{5.18}
f(v,r) = k - \frac{2m(v)}{r^{D-3}} + \frac{q^{2}(v)}{r^{2(D-3)}}
- \Lambda r^{2},
\eq
and the corresponding energy-momentum tensor can be written as
\bq
\lb{5.19}
{}^{(D)}T_{\mu\nu} = \mu l_{\mu}l_{\nu} + {}^{(D)}F_{\mu\alpha}
{}^{(D)}{F_{\nu}}^{\alpha}
- \frac{1}{4}{}^{(D)}F_{\alpha\beta}{}^{(D)}F^{\alpha\beta},
\eq
with
\bqn
\lb{5.20}
{}^{(D)}F_{\mu\nu} &=& \left(\frac{(D-2)(D-3)}{2}\right)^{1/2}
\frac{q(v)}{r^{D-2}} \left(\delta^{0}_{\mu}\delta^{1}_{\nu} -
\delta^{1}_{\mu}\delta^{0}_{\nu}\right),\nb\\
\mu &=& -  \frac{\epsilon(D-2)}{8\pi G_{D} r^{D-2}}
\left\{m'(v) - q(v)\frac{q'(v)}{r^{D-3}}\right\}.
\eqn

When $k = 1$ and $D = 4$ the above solutions reduce to the ones found in
\cite{WW99}. Lately, these solutions were generalized to high dimensional
spherical spacetimes  \cite{PD99}.

\section{Formation of Topological Black Holes  from Gravitational Collapse of a
Type II Fluid}

\renewcommand{\theequation}{6.\arabic{equation}}
\setcounter{equation}{0}

To consider gravitational collapse of the Type $II$ fluid found in the last section
let us choose $\epsilon = -1$. Then,  metric (\ref{5.2}) with $\psi = 0$
becomes
\bq
\lb{6.1}
ds^{2} = - f(v,r) dv^{2} + 2 dvdr
+ r^{2}h_{AB}\left(x^{C}\right)dx^{A}dx^{B}.
\eq
To the present purpose, let us consider the solutions given by
Eqs.(\ref{5.18}) - (\ref{5.20}) with
\bqn
\lb{6.2}
m(v) &=& \cases{\lambda {v_{0}}^{D-3}, & $ v \ge v_{0}$,\cr
\lambda v^{D-3}, & $ 0 \le v \le v_{0}$,\cr
0, & $ v< 0$,\cr}\nb\\
q(v) &=& \cases{\delta {v_{0}}^{D-3}, & $ v \ge v_{0}$,\cr
\delta v^{D-3}, & $ 0 \le v \le v_{0}$,\cr
0, & $ v< 0$,\cr}
\eqn
where  $\lambda$ and $\delta$ are arbitrary real constants, subject to $\lambda > 0$.
 Clearly, the solutions represent a charged null dust
fluid moving on a de Sitter or anti-de Sitter background in the region
$0 \le v \le v_{0}$, depending on the sign of $\Lambda$. When $\Lambda = 0$
the spacetime is self-similar, while when $\Lambda \not= 0$
it is only asymptotically self-similar, $(v, r) \rightarrow (0, 0)$.
The corresponding energy-momentum tensor is given by Eq.(\ref{5.19}) with
\bqn
\lb{6.3}
{}^{(D)}F_{\mu\nu} &=& \cases{
\left(\frac{(D-2)(D-3)}{2}\right)^{1/2}
\frac{\delta {v_{0}}^{D-3}}{r^{D-2}} \left(\delta^{0}_{\mu}\delta^{1}_{\nu} -
\delta^{1}_{\mu}\delta^{0}_{\nu}\right), & $    v > v_{0}$,\cr
\left(\frac{(D-2)(D-3)}{2}\right)^{1/2}
\frac{\delta y^{D-3}}{r} \left(\delta^{0}_{\mu}\delta^{1}_{\nu} -
\delta^{1}_{\mu}\delta^{0}_{\nu}\right), & $ 0 \le v \le v_{0}$,\cr
\;\;\;\;\;\;\;\;\;\;\;\; 0, & $v < 0$, \cr}\nb\\
\mu &=& \cases{\frac{(D-2)(D-3)\delta^{2}y^{D-4}}{8\pi G_{D} r^{2}}
\left({y_{c}}^{D-3} -  y^{D-3}\right),  & $  0 \le v \le v_{0}$,\cr
\;\;\;\;\;\;\;\;\;\;\;\; 0, & otherwise,\cr}
\eqn
where
\bq
\lb{6.4}
y \equiv \frac{v}{r}, \;\;\; y_{c} \equiv
\left(\frac{\lambda}{\delta^{2}}\right)^{1/(D-3)}.
\eq
From Eqs.(\ref{5.6a})-(\ref{5.6d}) we can see that the Ricci
tensor contains only the first order derivatives of $f$ with respect to $v$.
Thus, there would be no matter shell to appear on a hypersurface $v = Const.$,
as long as $f(v,r)$ is continuous across  this surface. Clearly, this is
the case for the choice of Eq.(\ref{6.2})
crossing the hypersurfaces $v = 0$ and $v = v_{0}$.

When $k = 1,\; \Lambda = 0 = \delta$, and $X^{D-2} = S^{D-2}$,
the corresponding solution is the Vaidya
solution in $D$-dimensional spherically symmetric spacetime \cite{PD99}, and was
studied independently in  \cite{GD01,Roch}.
In particular, da Rocha showed that when $\lambda > \lambda_{c}$ the collapse
always forms black holes, and when $\lambda < \lambda_{c}$ the collapse
always forms naked singularities, where
\bq
\lb{6.4a}
\lambda_{c} \equiv \frac{(D-3)^{D-3}}{\left[2(D-2)\right]^{D-2}}.
\eq
This result can be easily generalized to the case where $\Lambda
\not= 0,\; \delta = 0$ and
$X^{D-2}$ has a topology rather than $S^{D-2}$.
As a matter of fact, the case with  $k = 1,\; \Lambda > 0,\; \delta = 0$
and $X^{D-2} = S^{D-2}$ was already studied
in \cite{GD02}, from which we can see that Eq.(\ref{6.4}) is also valid
for any   $\Lambda$ and other topologies for $X^{D-2}$ but still with
$k = 1,\; \delta =0$.

When $k =1, \; \Lambda = 0,\; \delta \not= 0$ and $D = 4$,
Lake and Zannias found that the collapse
can form either black holes or naked singularities \cite{LZ91}. Ghosh generalized
these results to the case where $\Lambda > 0$ \cite{Ghosh02}. From the analysis given in
these two articles one can see that the cosmological constant actually has no
effects on the final state of the collapse. Therefore, the results obtained by
Lake and Zannias are actually valid for any $\Lambda$. In addtion, following their
analysis it is not difficult to be convinced that {\em gravitational collapse of
a charged Type $II$ fluid in a higher dimensional de Sitter ($\Lambda > 0$) or anti-de
Sitter ($\Lambda < 0$) spacetime with $k = 1$ can form either black holes or naked
singularities, depending on the choice  of $\lambda$ and $\delta$}.

Therefore, in the following we shall consider only the cases where $k = 0,
\; -1$. Let us first consider the case $\delta = 0$, i.e., the collapse of a
neutral null dust fluid. In this case from Eq.(\ref{5.18}) we can see
that to have $f(v,r)$ be
positive at least in some regions of the spacetime, we have to assume that
$\Lambda < 0$. Then, Eq.(\ref{5.18}) can be written as
\bqn
\lb{6.5}
f(v,r) &=& 2{\lambda}\left[\frac{|\Lambda|}{2\lambda}
\left(r^{2} - {r_{0}}^{2}\right) - y^{D-3}\right],\nb\\
r_{0} &\equiv& \left|\frac{k}{\Lambda}\right|^{1/2},
\eqn
while Eq.(\ref{A.10}) yields,
\bqn
\lb{6.6}
\theta_{l}\theta_{n} &=& \frac{2{\lambda}FG}{r^{D-1}}
\left(v^{D-3} - \frac{|\Lambda|}{2\lambda}
r^{D-3}\left(r^{2} - {r_{0}}^{2}\right)\right)\nb\\
&=& \cases{< 0, & $v < v_{AH}(r)$,\cr
=0, & $v = v_{AH}(r)$,\cr
> 0, & $v > v_{AH}(r)$,\cr}
\eqn
where
\bq
\lb{6.7}
v_{AH}(r) \equiv r \left\{\frac{|\Lambda|}{2\lambda}
\left(r^{2} - {r_{0}}^{2}\right)\right\}^{1/(D-3)}.
\eq
Thus, in the present case the ($D-2$)-surfaces of constant $v$ and $r$
are trapped in the region  $v > v_{AH}(r)$. The hypersurface
$v = v_{AH}(r)$ represents an apparent horizon [cf. Fig. 12].
Since the spacetime singularity at $r = 0$ starts to be formed only
at the moment $v = 0$, from Fig. 12 we can see that this singularity
is always covered by the apparent horizon. This can be seen further
by studying ``outgoing" null geodesics, which are given by
\bq
\lb{6.8}
\frac{dv}{dy} = \frac{2}{|\Lambda|r^{2} - |k| - 2\lambda y^{D-3}}.
\eq
At the moment $v = 0$  if there exist out-going null geodesics
from the point $(v, r) = (0, 0)$, we can see that the singularity
formed at that moment will be at least locally naked. The existence
of such null geodesics is characterized by the existence of positive
roots of the equation \cite{Joshi},
\bq
\lb{6.9}
2{\lambda} {y_{0}}^{D-2} + |k| y_{0} + 2 = 0,
\eq
where
\bq
\lb{6.10}
y_{0} \equiv \lim_{ v, r \rightarrow 0 }{\frac{v}{r}}
= \lim_{v, r \rightarrow 0}{\frac{dv}{dr}}.
\eq
Since all the coefficients of Eq.(\ref{6.9}) are positive, no
positive roots exist. As a result, the collapse will always form
black holes with non-trivial topology. This generalizes the results
obtained in four-dimensional case \cite{SM97} to the one with any dimensions.

 \begin{figure}[htbp]
 \begin{center}
 \label{fig11}
 \leavevmode
  \epsfig{file=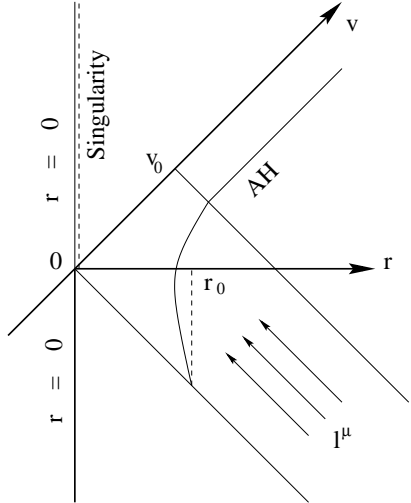,width=0.3\textwidth,angle=0}
\caption{Collapsing Type $II$ fluid described by Eq.(\ref{6.2})
 in the text with $k \le 0, \; \delta = 0$, and $\Lambda < 0$.
The curved line $AH$ represent an apparent horizon, and the
collapse always forms black holes. }
 \end{center}
 \end{figure}

When $\delta \not = 0$, Eq.(\ref{6.9}) should be replaced by
\bq
\lb{6.11}
 \delta^{2}{y_{0}}^{2D-5} - 2\lambda{y_{0}}^{D-2}
- |k|y_{0} - 2 = 0.
\eq
It is important to note that in the present case the charged fluid satisfies
the weak energy condition only in the region $y \le y_{c}$, as we can see from
Eq.(\ref{6.3}). On the hypersurface $y = y_{c}$ the energy density becomes
zero. Afterwards, the Lorentz force will push the fluid particles
 to move outwards \cite{Ori91}. As a result, the particles actually
cannot enter into the region $y > y_{c}$ [cf. Fig. 13]. Now to see if the spacetime
singularity formed at $(v, r) = (0, 0)$ is naked or not, one needs not only to
show that a positive root of Eq.(\ref{6.11}) exists, but also to show that
the outgoing null geodesics fall inside the region $y \le y_{c}$, where
the solution is actually valid.

 \begin{figure}[htbp]
 \begin{center}
 \label{fig12}
 \leavevmode
  \epsfig{file=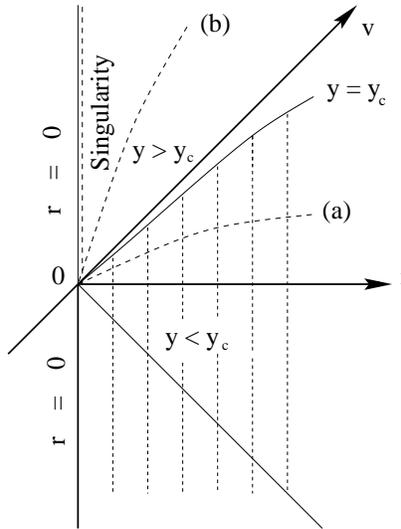,width=0.3\textwidth,angle=0}
\caption{Null geodesics for collapsing Type $II$ fluid described by Eq.(\ref{6.2})
with $\delta \not= 0$. The dashed line $(a)$ represents an outgoing null
geodesic with $y_{0}^{s} < y_{c}$, while the one $(b)$ an outgoing null
geodesic with $y_{0}^{s} > y_{c}$, where $y^{s}_{0}$ denotes the positive
root of Eq.(\ref{6.11}) and $y_{c}$ is defined by Eq.(\ref{6.4}). }
 \end{center}
 \end{figure}

From Fig. 13 we can see that if an outgoing null geodesic falls into the region
$y < y_{c}$ we must have $y_{0}^{s} < y_{c}$, and  if an outgoing null geodesic
falls into the region $y > y_{c}$, we must have $y_{0}^{s} > y_{c}$, where
$y_{0}^{s}$ is the smallest positive root of Eq.(\ref{6.11}).
Since in the region $y > y_{c}$ we have $\mu < 0$ and in a realistic model
this region should be replaced by an out-going charged dust fluid. Thus, in the
latter case the singularity at $(v, r) = (0, 0)$ should not be considered as
naked, but in the former case it is. Therefore, to see if the singularity is naked or
not now reduces to find out if $y_{0}^{s} < y_{c}$ or $y_{0}^{s} > y_{c}$.
To this end, let us first consider the case $k = 0$, and define the function
$G\left(y_{0}\right)$ by
\bq
\lb{6.12}
G\left(y_{0}\right) \equiv \delta^{2}{y_{0}}^{2D-5} - 2\lambda{y_{0}}^{D-2}
 - 2, \;\;\; (k = 0).
\eq
From this expression we can see that $G'\left(y_{0}\right) = 0$ has two roots,
$y^{\pm}_{0}$, given by
\bq
\lb{6.13}
y_{0}^{-} = 0,\;\;\;
y_{0}^{+} = \left(\frac{2\lambda(D-2)}{\delta^{2}(2D-5)}\right)^{1/(D-3)},
\eq
and
\bqn
\lb{6.14}
G\left(y^{-}_{0}\right) & = & -2,\nb\\
G\left(y^{+}_{0}\right) & = & - \frac{4\lambda^{2}(D-2)(D-3)}
{\delta^{2}(2D-5)^{2}}{y^{+}_{0}}^{1/(D-3)} - 2 < -2,\nb\\
G\left(y_{c}\right) &=& - \frac{\lambda^{2}}{\delta^{2}}
  \left(\frac{\lambda}{\delta^{2}}\right)^{1/(D-3)} - 2 < -2.
\eqn
Then, the curve of $G\left(y_{0}\right)$ vs $y_{0}$ must be that given by
the line (a) in Fig. 14, from which we can see that
\bq
\lb{6.15}
y^{s}_{0} > y_{c}.
\eq
That is, in the present case there does not exist outgoing null geodesics
in the region $y \le y_{c}$. So, the singularity formed at the point
$(v, r) = (0, 0)$ is not naked.

When $k = -1$, setting
\bq
\lb{6.16}
F\left(y_{0}\right) \equiv \delta^{2}{y_{0}}^{2D-5} - 2\lambda{y_{0}}^{D-2}
- y_{0} - 2,\;\;\; (k = -1),
\eq
we find that
\bq
\lb{6.17}
G\left(y_{0}\right) - F\left(y_{0}\right) = y_{0},
\eq
and the curve of $F\left(y_{0}\right)$ vs $y_{0}$ must be given by the dashed
line (b) in  Fig. 14. From there we can see clearly that Eq.(\ref{6.15})
also holds for $k = -1$.

 \begin{figure}[htbp]
 \begin{center}
 \label{fig13}
 \leavevmode
  \epsfig{file=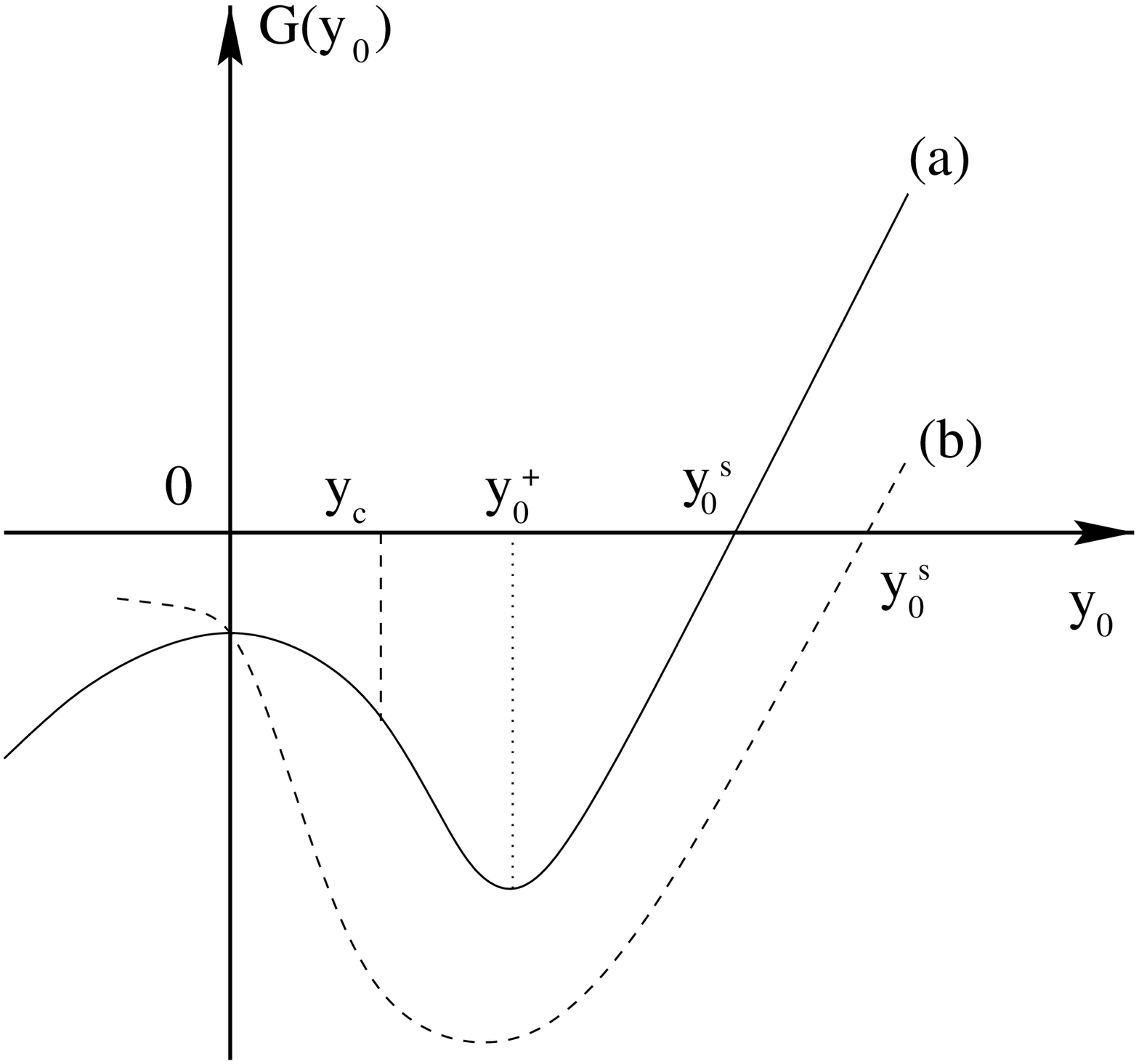,width=0.3\textwidth,angle=0}
\caption{(a) The function of $G\left(y_{0}\right)$ defined by Eq.(\ref{6.12})
for $k = 0$. (b) The function of $F\left(y_{0}\right)$ defined by Eq.(\ref{6.16})
for $k = -1$.  }
 \end{center}
 \end{figure}

Therefore it is concluded that the {\em gravitational collapse of a charged null
dust fluid given by Eqs.(\ref{5.18}), (\ref{6.1}) and (\ref{6.2}) with $k = 0$ or
$k = -1$  always forms black holes}. This is in contrast to the results obtained
in \cite{Ghosh02}. The reason is that  in \cite{Ghosh02} the author didn't consider
the energy condition of the charged null dust fluid.

\section{Conclusions}

In this paper, we have studied topological black holes and their
formation from gravitational collapse of a Type $II$ fluid in
$D$-dimensional spacetimes described by the metric Eq.(\ref{2.1}). After presenting a
general  $R^{2}\times X^{D-2}$ decomposition in Sec. $II$, we have
re-derived the charged solutions coupled with a cosmological
constant, given by Eq.(\ref{3.16}), in Sec. $III$. From their derivations
we can see that these solutions are the most general ones for the spacetimes
described by Eq.(\ref{2.1}).

In Sec. $IV$, we have systematically studied the global structure
of the spacetime for $k_{D} = 0$ and $D = 4$, that is, four-dimensional
spacetimes with flat topology of the $X^{2}$ sector, which can be
 a two-dimensional plane,  cylinder,   M\"obius band,   torus,
or  Klein bottle, depending on how to  identify
the two coordinates  in $X^{2}$ \cite{Wolf67}.
All the corresponding Penrose diagrams
have been given. In particular, it has been found that the solution
with $\Lambda < 0$ and $q = 0$ has a black hole structure quite similar
to the Schwarzschild black hole [cf. Fig. 9].
In the case where $\Lambda < 0,\; bq \not= 0$, the global structure
of the corresponding spacetime is quite similar to the Reissner-Nordstr\"om
solution, including the extreme case [cf. Figs. 10 and 11].

In Sec. $V$, all the solutions of a Type $II$ fluid have been found, while
in Sec. $VI$ the gravitational collapse of such a fluid has been studied.
When $k = 1$ the collapse in a de Sitter or anti-de Sitter background
can form either black holes or naked
singularities, but when $k = 0$ or $k = -1$ it always form black holes.
Therefore, all the black hole solutions with different topologies found in
Sec. $IV$ can be realized from the gravitational collapse of a type $II$
fluid in D-dimensional spacetimes.

\section*{Acknowledgments}

One of the authors (AW) would like to thank   Nigel Goldenfeld  for his
valuable discussions  on   gravitational collapse. He would also like to express his
gratitude to the Department of Physics, UIUC,
for hospitality. The financial assistance from CAPES (AW),  CNPq (NOS) and
FAPERJ (MFAdaS) is gratefully acknowledged.

\section*{Appendix: Trapped Surfaces and Apparent Horizons}
\lb{APP}
\renewcommand{\theequation}{A.\arabic{equation}}
\setcounter{equation}{0}

The concept of {\em trapped surface} was originally from Penrose \cite{Pen68},
who defined it as a compact spatial two-surface $S$ in a four-dimensional spacetime,
on which $\left.\theta_{+}\theta_{-}\right|_{S}
> 0$, where $\theta_{\pm}$ denote the expansions in the future-pointing null
directions orthogonal to $S$, and the spacetime is assumed to be time-orientable,
so that ``future" and ``past" can be assigned consistently. One may then define
a past trapped surface by $\left.\theta_{\pm}\right|_{S} > 0$, and a future
trapped surface by $\left.\theta_{\pm}\right|_{S} < 0$.

Recently, Hayward \cite{Hay00} generalized the above
definition to the four-dimensional cylindrical spacetimes
where the two-surface $S$ is not compact but  an infinitely long cylinder,
and call it trapped, marginal or untrapped, according to where $\alpha_{,\mu}$ is
timelike, null or spacelike, where $\alpha$ is  defined by
\bq
\lb{A.1}
\alpha \equiv \left(\left|\partial_{z} \cdot \partial_{z}\right| \cdot
\left|\partial_{\varphi}
\cdot \partial_{\varphi}\right|\right)^{1/2},
\eq
with $\xi_{(z)} = \partial_{z}$ and  $\xi_{(\varphi)} = \partial_{\varphi}$
being the two killing vector in the cylindrical spacetimes.
Lately, one of the present authors showed that Hayward's definition is
consistent with the one defined by the expansions of the two null directions
orthogonal to the two-cylinder \cite{Wang03}.

In this Appendix, we shall generalize the above definition to the D-dimensional
case described by the metric (\ref{5.2}). Now let us consider the (D-2)-surface,
$S$, of constant $v$ and $r$. When $k = 1$ it could be a $(D-2)$-sphere, although
other topologies can also exist.
When $k \not= 1$ it can be compact or non-compact,  depending on how to
identify the coordinates in $S$.   For more
details, we refer readers to \cite{Wolf67,ABHP} and references therein.

To calculate the expansions of   the two null directions orthogonal to the
$(D-2)$-surface $S$ of constant $x^{a}$ in the metric (\ref{5.2}), it is
found convenient to introduce two null coordinates
$\bar{v}$ and $\bar{u}$ via the relations
\bqn
\lb{A.2a}
d\bar{u} &=& G(v,r)\left(fe^{\psi} dv + 2 \epsilon dr\right),\nb\\
d\bar{v} &=& F(v) dv,
\eqn
or inversely
\bqn
\lb{A.2b}
dr &=& \frac{\epsilon}{2}\left(\frac{1}{G} d\bar{u}
- \frac{1}{F} fe^{\psi} d\bar{v}\right),\nb\\
d{v} &=& \frac{1}{F} d\bar{v},
\eqn
where $G(v,r)$ is determined by the integrability condition $\bar{u}_{,vr}
= \bar{u}_{,rv}$, and $F(v)$ is an arbitrary function of $v$ only.
Without loss of generality we shall assume that they are all strictly
positive,
\bq
\lb{A.3a}
F(v) > 0,\;\;\;\;\; G(v, r) > 0.
\eq
Then, in terms of $\bar{u}$ and $\bar{v}$ the metric (\ref{5.2}) takes
the form
\bq
\lb{A.3}
ds^{2} = - 2 e^{2\sigma} d\bar{u} d\bar{u}
+ r^{2}h_{AB}\left(x^{C}\right)dx^{A}dx^{B},
\eq
where $\sigma$ and $r$ are now the functions of $\bar{u}$ and $\bar{v}$
via Eq.(\ref{A.2a}) and
\bq
\lb{A.4}
\sigma \equiv \frac{1}{2} \left[\psi - \ln\left(2FG\right)\right].
\eq
Clearly, the metric (\ref{A.3}) is invariant under the coordinate transformations
\bq
\lb{A.5}
\bar{u} = \bar{u}(\tilde{u}),\;\;\;\;
\bar{v} = \bar{v}(\tilde{v}).
\eq
Using this gauge freedom, we shall  assume that metric (\ref{A.3})
is free of coordinate singularities. Then, it can be shown that
\bqn
\lb{A.5a}
{}^{(2)}R_{ab} &=& - 2\sigma_{,\bar{u}\bar{v}} \left(\delta^{0}_{a}\delta^{1}_{b}
+ \delta^{1}_{a}\delta^{0}_{b}\right),\nb\\
\nabla_{a}\nabla_{b}r &=& r_{,ab} - 2\left(\sigma_{,\bar{u}} r_{,\bar{u}}
\delta^{0}_{a}\delta^{0}_{b}
+ \sigma_{,\bar{v}} r_{,\bar{v}} \delta^{1}_{a}\delta^{1}_{b}\right).
\eqn
Substituting Eq.(\ref{A.5a})  into Eq.(\ref{2.7})
we find that
\bqn
\lb{A.5ba}
{}^{(D)}R_{\bar{u}\bar{u}} &=& - \frac{D-2}{r}\left(r_{,\bar{u}\bar{u}}
- 2 \sigma_{,\bar{u}} r_{,\bar{u}}\right),\nb\\
{}^{(D)}R_{\bar{u}\bar{v}} &=& - 2\sigma_{,\bar{u}\bar{v}}
- \frac{D-2}{r}r_{,\bar{u}\bar{v}},\nb\\
{}^{(D)}R_{\bar{v}\bar{v}} &=& - \frac{D-2}{r}\left(r_{,\bar{v}\bar{v}}
- 2 \sigma_{,\bar{v}} r_{,\bar{v}}\right),\nb\\
{}^{(D)}R_{AB} &=& \left\{k_{D} + 2 e^{-2\sigma} \left[r r_{,\bar{u}\bar{v}}
+   \left(D-3\right)r_{,\bar{u}} r_{,\bar{v}}\right]\right\}h_{AB}.
\eqn

On the other hand, introducing two null vectors
$l_{\mu}$ and $n_{\mu}$
by
\bq
\lb{A.6}
 l_{\lambda} \equiv \frac{\partial \bar{u}}{\partial x^{\lambda}} =
 \delta^{\bar{u}}_{\lambda},\;\;\;
 n_{\lambda} \equiv \frac{\partial \bar{v}}{\partial x^{\lambda}} =
 \delta^{\bar{v}}_{\lambda},
\eq
one can see that the two null vectors are future directed, and orthogonal
to the ($D-2$)-surface, $S$, of constant $\bar{u}$ and $\bar{v}$
(or constant of $v$ and $r$). In addition, each of them
defines an affinely parametrized null geodesic congruence, since now
the following holds
\bqn
 \lb{A.7}
\frac{D}{D\lambda}l^{\mu} &=& {l^{\mu}}_{;\nu} l^{\nu} = 0,\nb\\
\frac{D}{D\delta}n^{\mu} &=& {n^{\mu}}_{;\nu} n^{\nu} = 0,
\eqn
where a semicolon ``;" denotes the covariant derivative with respect to
$g_{\mu\nu}$, and $\lambda$ and
$\delta$ the affine parameters along the null rays defined, respectively,
by $l^{\mu}$ and $n^{\mu}$ (It should be noted that the symbol $\delta$ used
here should not be confused with the one used in Eq.(\ref{6.2}) for the charge
density.). In particular,
$l^{\mu}$ defines the one moving along the null
hypersurfaces $\bar{u} =  Const.$, while $n^{\mu}$ defines the one
moving along the null hypersurfaces $\bar{v} =  Const.$
Then, the expansions of these null geodesics are defined by \cite{HE73},
\bqn
  \lb{A.8}
\theta_{l} &\equiv& - \frac{1}{D} g^{\alpha\beta} l_{\alpha;\beta}
   =  e^{-2\sigma}\frac{r_{,\bar{v}}}{r},\nb\\
\theta_{n} &\equiv& - \frac{1}{D} g^{\alpha\beta} n_{\alpha;\beta}
  =  e^{-2\sigma}\frac{r_{,\bar{u}}}{r}.
\eqn
Thus, we have
\bq
\lb{A.8aa}
 \theta_{l} \theta_{n}
   =  e^{-4\sigma}\frac{r_{,\bar{v}}r_{,\bar{u}}}{r^{2}}
   = - \frac{1}{2r^{2}}e^{-2\sigma} \; r_{,\alpha} r^{,\alpha}.
\eq

It should be noted that the affine parameter
$\lambda$ (or $\delta$) is unique only up to a function $f^{-1}(\bar{u})$
(or $g^{-1}(\bar{v})$), which is a constant along each curve $\bar{u} = Const.$
 (or $\bar{v} = Const.$) \cite{HE73}.
In fact, $\bar{\lambda} = \lambda/f(\bar{u})$ ($\bar{\delta} = \delta/g(\bar{v})$)
is  another affine parameter and the corresponding tangent vectors
are
\bq
\lb{A.6b}
\bar{l}_{\mu} = f(\bar{u})\delta^{\bar{u}}_{\mu},\;\;\;\;
\bar{n}_{\mu} = g(\bar{v})\delta^{\bar{v}}_{\mu},
\eq
and the corresponding   expansions
are given by
\bq
\lb{A.8b}
\bar{\theta}_{\bar{l}} = f(\bar{u})\theta_{l},\;\;\;
\bar{\theta}_{\bar{n}} = g(\bar{v})\theta_{n}.
\eq
However, since along each curve
$\bar{u} = Const.$ (or $\bar{v} = Const.$) the function $f(\bar{u})$ (or $g(\bar{v})$)
is constant, this does not affect our definition of trapped surfaces in terms
of the expansions. Thus, without loss of generality,
in the following we consider only the expressions given by Eq.(\ref{A.8}).

Once we have the expansions, following Penrose we can define that {\em a
($D-2$)-surface, $S$, of constant $\bar{u}$ and $\bar{v}$ is trapped, marginally trapped,
or untrapped, according to whether $\theta_{l}\theta_{n} > 0$,
$\; \theta_{l}\theta_{n} = 0$, or $\theta_{l}\theta_{n} < 0$.
An apparent horizon, or trapping horizon in Hayward's terminology \cite{Hay00},
is defined as a hypersurface foliated by marginally trapped
surfaces}.

Since $e^{-2\sigma}$ is regular, except at some points or surfaces on
which the spacetime is singular, from Eq.(\ref{A.8aa}) we can see that
trapped, marginally trapped, or untrapped surfaces can be also defined
according to whether $r_{,\alpha}$ is timelike, null, or spacelike.

On the other hand, from Eq.(\ref{A.2b}) we find that
\bq
\lb{A.9}
\frac{\partial r}{\partial \bar{u}} = \frac{\epsilon}{2G},\;\;\;\;
\frac{\partial r}{\partial \bar{v}} = - \frac{{\epsilon}f}{2F}e^{\psi}.
\eq
Inserting the above expressions into Eq.(\ref{A.8}) we obtain
\bqn
\lb{A.10}
\theta_{l} &\equiv& - \frac{1}{D} g^{\alpha\beta} l_{\alpha;\beta}
   =  e^{-2\sigma}\frac{r_{,\bar{v}}}{r},\nb\\
   &=& - \frac{{\epsilon}f}{2rF}e^{\psi-2\sigma},\nb\\
\theta_{n} &\equiv& - \frac{1}{D} g^{\alpha\beta} n_{\alpha;\beta}
  =  e^{-2\sigma}\frac{r_{,\bar{u}}}{r}\nb\\
  & =& \frac{{\epsilon}}{2rG}e^{-2\sigma},\nb\\
  \theta_{l}\theta_{n} &=& - \frac{FG}{r^{2}}f e^{-\psi}.
\eqn

\end{document}